\documentclass[iopams]{iopart}
\usepackage{geometry}
\usepackage{graphics}
\usepackage{graphicx}
\usepackage{amssymb}
\usepackage{enumerate}
\usepackage{color}
\usepackage{subfig}

\DeclareMathAlphabet\mathbit
    \encodingdefault\rmdefault\bfdefault\itdefault
\DeclareOldFontCommand{\bi}{\normalfont\bfseries\itshape}{\mathbit}

\newcommand{\be}{\begin{equation}}
\newcommand{\ee}{\end{equation}}

\def\fakebold#1{\relax\ifvmode\leavevmode\fi%
\ifmmode%
\setbox0=\hbox{$#1$}%
\else%
\setbox0=\hbox{#1}%
\fi%
\kern-.02em\copy0 \kern-\wd0%
\kern .04em\copy0 \kern-\wd0%
\kern-.0125em\raise.02em\box0%
}%

\begin{document}
\ifx\aligned\undefined

\makeatletter
\def\aligned{{\ifnum0=`}\fi\vcenter\bgroup\let\\\cr
\halign\bgroup&\hfil$\displaystyle{}##{}$&$\displaystyle{}##{}$\hfil\cr}
\def\endaligned{\crcr\egroup\egroup\ifnum0=`{\fi}}

\def\align{\par
\bigskip
{\ifnum0=`}\fi
\let\\\cr
\halign to \textwidth\bgroup
\refstepcounter{equation}%
\global\let\@alignlab\@currentlabel
\vrule \@height \dimexpr\ht\strutbox+3pt\relax
       \@depth  \dimexpr\dp\strutbox+1pt\relax
       \@width \z@
\hbox to \textwidth{\hfill(\theequation)}\kern-\textwidth
\tabskip\fill
\hfil$\displaystyle{}##{}$&%
\let\@currentlabel\@alignlab$\displaystyle{}##{}$\hfil&%
\let\@currentlabel\@alignlab\hfil$\displaystyle{}##{}$&%
\let\@currentlabel\@alignlab$\displaystyle{}##{}$\hfil\cr}
\def\endalign{\crcr\egroup\ifnum0=`{\fi}\par\bigskip}

\newcounter{parentequation}
\newenvironment{subequations}{%
  \refstepcounter{equation}%
  \edef\theparentequation{\theequation}%
  \setcounter{parentequation}{\value{equation}}%
  \setcounter{equation}{0}%
  \def\theequation{\theparentequation\alph{equation}}%
  \ignorespaces
}{%
  \setcounter{equation}{\value{parentequation}}%
  \ignorespacesafterend
}

\def\cases{{\ifnum0=`}\fi\left\{\array{lll}}
\def\endcases{\endarray\right.\ifnum0=`{\fi}}

\makeatother
\fi

\title[asymptotic~approximants]{Asymptotic Approximant for the Falkner-Skan Boundary-Layer equation}
\author{Elizabeth R. Belden$^1$, Zachary A. Dickman$^1$, Steven J. Weinstein$^{1}$, Alex D. Archibee$^{2}$, Ethan Burroughs$^{3}$, Nathaniel S. Barlow$^{3}$}
\address{$^1$ Department of Chemical Engineering, Rochester Institute of Technology, Rochester, NY 14623} 
\address{$^2$ Department of Mechanical Engineering, Rochester Institute of Technology, Rochester, NY 14623} 
\address{$^3$ School of Mathematical Sciences, Rochester Institute of Technology, Rochester, NY 14623} 
\ead{nsbsma@rit.edu}

\begin{abstract} 
We demonstrate that the asymptotic approximant applied to the Blasius boundary layer flow over a flat plat (Barlow \etal, 2017 \textit{Q. J. Mech. Appl. Math.}, \textbf{70} (1): 21-48) yields accurate analytic closed-form solutions to the Falkner-Skan boundary layer equation for flow over a wedge having angle $\beta\pi/2$ to the horizontal.   A wide range of wedge angles satisfying $\beta\in[-0.198837735, 1]$ are considered, and the  previously established non-unique solutions for $\beta<0$ having positive and negative shear rates along the wedge are accurately represented.  The approximant is used to determine the singularities in the complex plane that prescribe the radius of convergence of the power series solution to the Falkner-Skan equation. An attractive feature of the approximant is that it may be constructed quickly by recursion compared with traditional Padé approximants that require a matrix inversion.  The accuracy of the approximant is verified by numerical solutions, and benchmark numerical values are obtained that characterize the asymptotic behavior of the Falkner-Skan solution at large distances from the wedge.  \end{abstract}

\section{Introduction \label{sec:intro}}
\begin{figure}[h!]
\begin{center}
\includegraphics[width=5.5in]{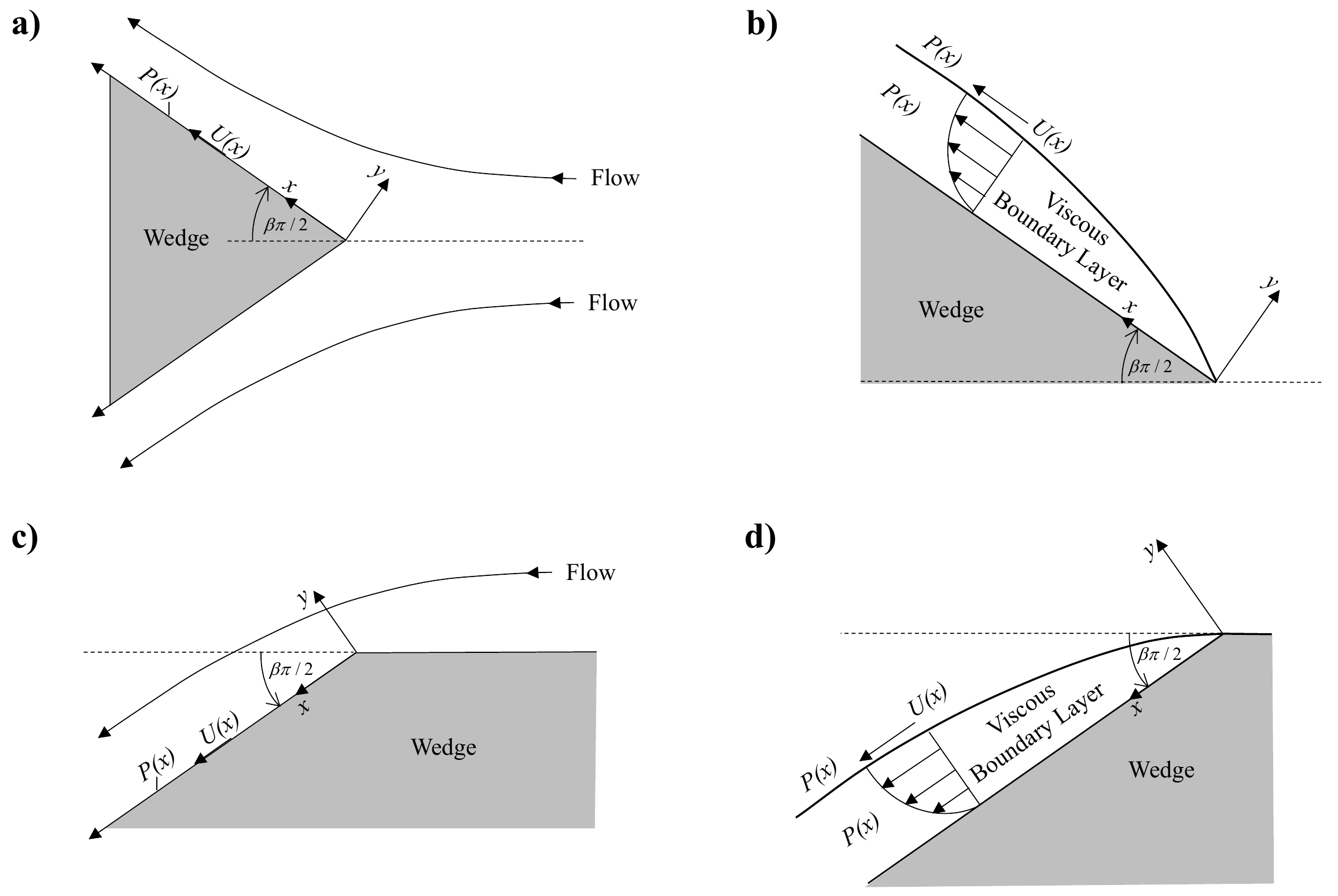}
\end{center}
\caption{Geometry of flow over a wedge showing potential flow and viscous boundary layer regions. a) Domain for the potential flow solution for $\beta>0$, from which the nonzero velocity $U(x)$ and associated Bernouli pressure $P(x)$ along $y=0$ is extracted. b) Domain for the thin viscous boundary layer (Falkner-Skan problem) that is driven by the potential flow solutions for $U(x)$ and $P(x)$. There are no pressure variations in the $y$-direction. On the length scale of figure a) the boundary layer cannot be observed. The analogous wedge configuration for $\beta<0$ are shown in Figures c) and d).}
\label{fig:schematic}
\end{figure}

The Falkner-Skan equation describes boundary layer flow over a wedge of angle $\beta\pi/2$ to the horizontal that is driven by an external pressure gradient predicted from potential flow (see Fig.~\ref{fig:schematic}). The equation also applies to regimes where the pressure gradient opposes the flow when $\beta<0$ (Fig.~\ref{fig:schematic}c,d) for which boundary layer separation may occur.  Through a similarity transform, the governing 2D boundary-layer equations can be written as a nonlinear ODE system for the dimensionless stream function, $f$, as a function of a similarity variable $\eta(x, y)$~\cite{Schlichting}:
\begin{eqnarray}
\nonumber
&f'''+ff''+\beta(1-f'^{2})=0\\
&f(0)=0, f'(0)=0, f'({\infty})=1.
\label{diffeq}
\end{eqnarray}
Once~(\ref{diffeq}) is solved for $f(\eta)$, the dimensional stream function, $\psi(x,y)$, may be extracted as~\cite{Schlichting} 
\begin{equation}
\eta=y\sqrt{\frac{m+1}{2}\frac{U}{\nu x}}
\label{eq:eta}
\end{equation}
\[\psi(x,y)=\sqrt{\frac{2\nu Ux}{m+1}}f(\eta)\]
where $x$ and $y$ are coordinates along and perpendicular to the wedge surface (see Fig.~\ref{fig:schematic}), $m=\beta/(2-\beta)$, $\nu$ is kinematic viscosity, and $U=U(x)$ is the velocity at the wedge surface determined from potential flow, as indicated in Fig.~\ref{fig:schematic}.  Although a power series solution can be found for~(\ref{diffeq}), it diverges at a finite value of $\eta$ for all physical values of wedge angle $\beta$. The lack of an exact analytical solution has necessitated several numerical~\cite{Cebeci,Laine,Fazio:1996,Asaithambi:1998,Motsa,Fazio:2013,Liu} and approximate analytical~\cite{Bararnia:2012,Yun:2012,Khidir:2015} solution approaches to the system~(\ref{diffeq}). Here, we implement the recent method of asymptotic approximants to analytically continue the power series solution, and thereby construct a highly accurate closed form solution to~(\ref{diffeq}). 

 Asymptotic approximants, originally developed for divergent series problems of thermodynamics~\cite{BarlowJCP,BarlowAIChE,Barlow2015}, may be constructed when asymptotic behaviors are known in two different regions of a domain; implementation details are given in~\cite{Barlow:2017}.  The method is a generalization of two well-known mathematical techniques: asymptotic matching and Pad\'e approximants~\cite{Bender}.  Whereas asymptotic matching leads to a single expression that combines two \textit{overlapping} asymptotic expansions, asymptotic approximants do not require that the two series overlap; if the two series diverge before overlap, an asymptotic approximant has the ability to analytically continue each series and accurately fill-in the gap between them.   In the same way as Pad\'e approximants, asymptotic approximants are constructed such that the series expansion of the approximant about a given point is the same as that of the true expansion about that point.   However, whereas Pad\'e approximants are restricted to rational functions and thus have a specific asymptotic behavior about a chosen expansion point, asymptotic approximants are tailor-made to have the correct behavior in both regions of the domain.  For example, asymptotic approximants are shown to accurately describe the light trajectory around a Kerr black hole~\cite{Beachley,Barlow:2017b} by incorporating the correct logarithmic behavior near the black hole; a Pad\'e is incapable of representing such behavior efficiently.  Asymptotic approximants are used to construct accurate solutions for boundary flows over a stationary flat plate (the Blasius problem) and for a flat plate moving through a stationary fluid (the Sakiadis problem)~\cite{Barlow:2017}.   The Blasius approximant uses a power series expansion about $\eta=0$ and the leading-order behavior as $\eta\to\infty$; the full asymptotic behavior in this latter regime is not needed to obtain a highly accurate solution.  The leading-order $\eta\to\infty$ behavior of Blasius is exactly the same as that for the Falkner-Skan problem~(\ref{diffeq}) for flow over a wedge.  Thus the asymptotic approximant used in~\cite{Barlow:2017} for the Blasius problem may be applied to the Falkner-Skan equation.  In this work, we demonstrate that this may be done with high accuracy. 


The paper is organized as follows.  In Section~\ref{sec:series}, the series solution of~(\ref{diffeq}) and numerically-obtained asymptotic properties are provided as a function of wedge angle.  The Blasius asymptotic approximant of~\cite{Barlow:2017} is applied to the Falkner-Skan equation in Section~\ref{sec:approximant}; it is validated against numerical solutions for key values of wedge angle.  The Blasius-type approximant is found to be accurate for positive and negative shear rates at the wedge surface over a significant range of wedge angles \--- the range of $\beta$ encompasses all physical configurations for which there is no boundary layer separation.  An alternative asymptotic approximant is introduced to handle non-monotonic solutions to the Falkner-Skan equation that occur over a small range of negative shear rates at the wedge surface for $\beta<0$.  In Section~\ref{sec:complex}, the Blasius-type approximant is used to extract the radius of convergence of the Falkner-Skan series as a function of wedge angle; to the authors' knowledge, this is the first time this dependence has been disclosed.    Concluding remarks are made in section~\ref{sec:conclusions}.  

\section{Asymptotic properties and series expansion of the Falkner-Skan equation \label{sec:series}}
\begin{figure}[!ht]
\begin{center}
\subfloat{(a) \includegraphics[width=3in]{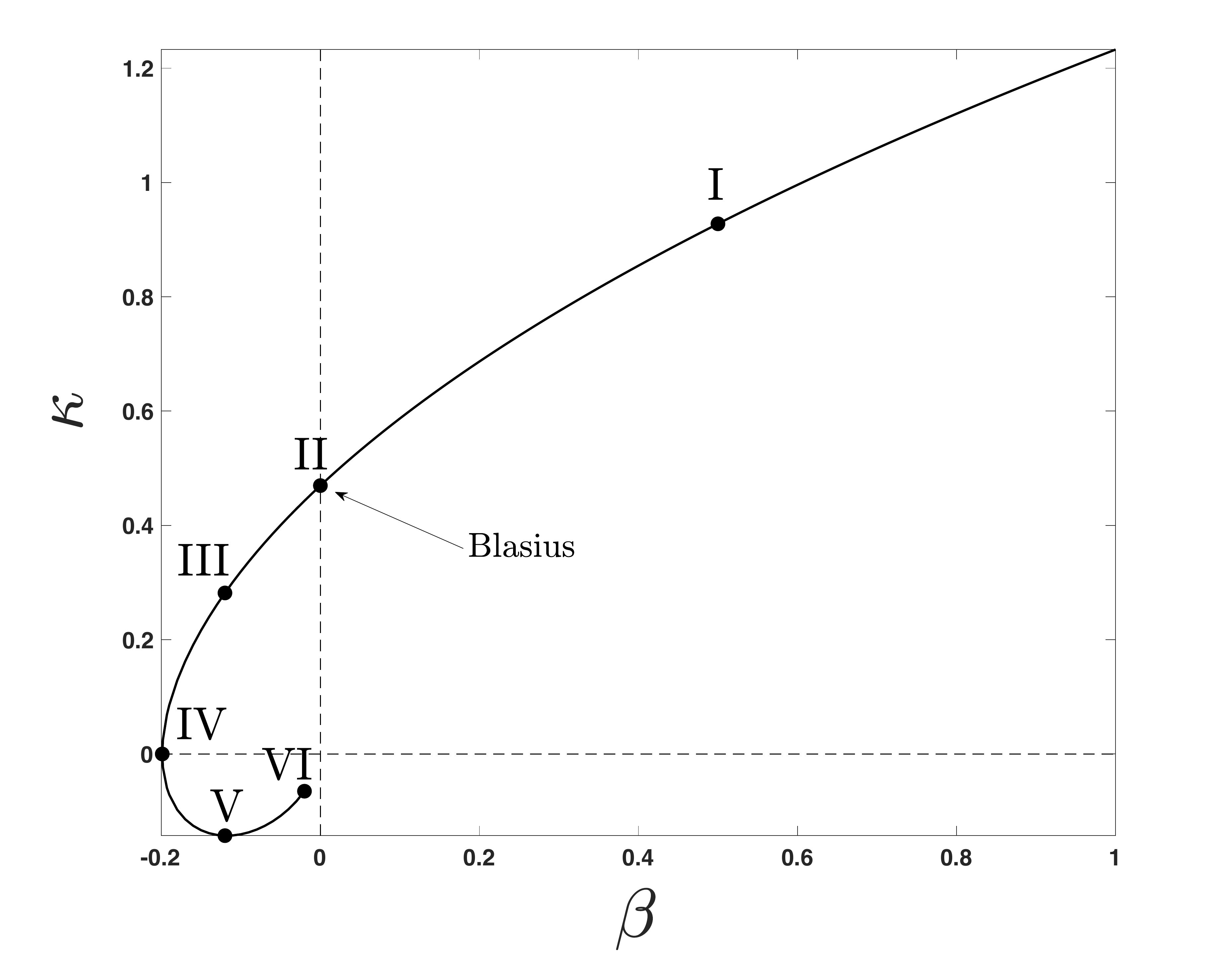}}
 \subfloat{(b) \includegraphics[width=3in]{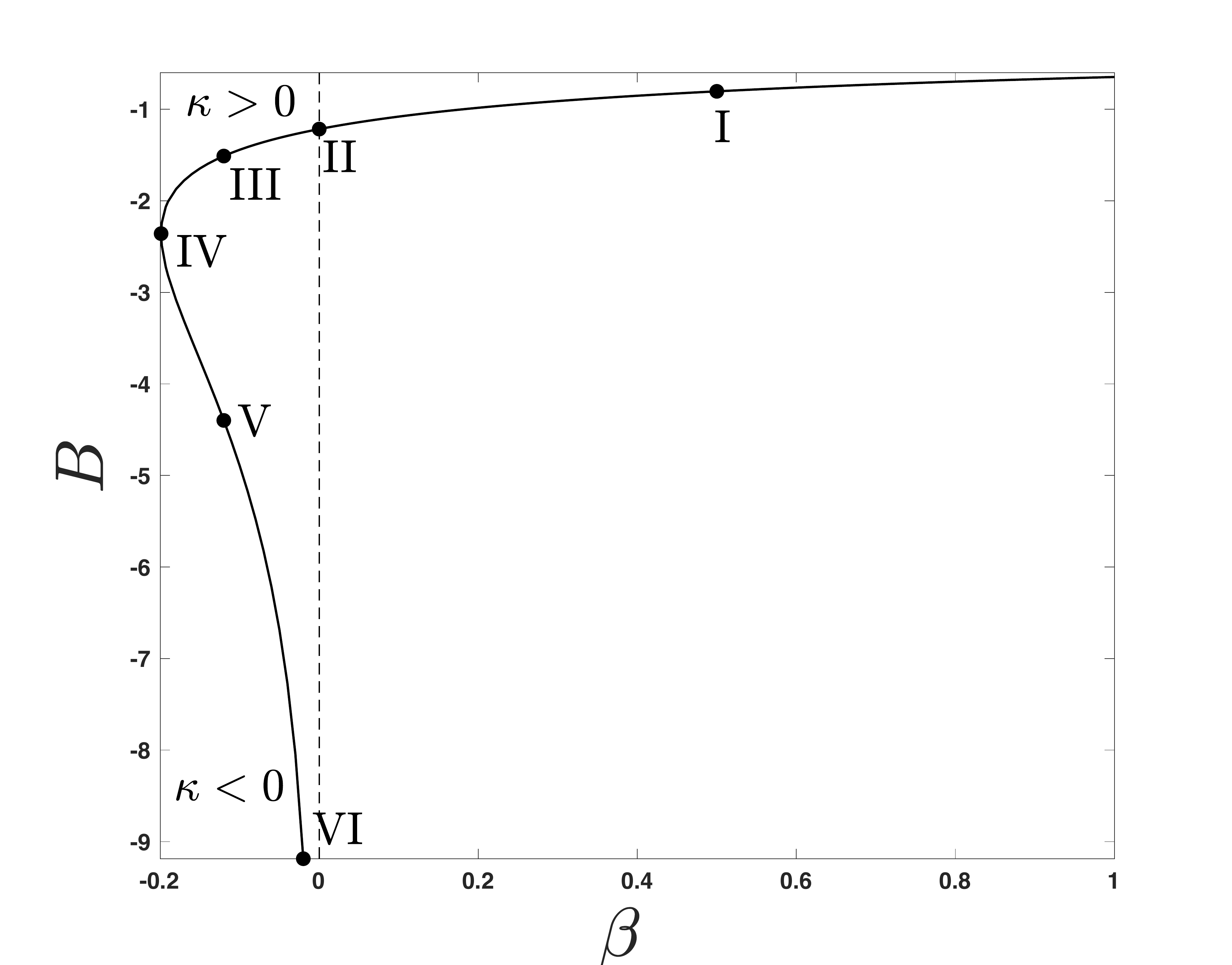}}
 \end{center}
\caption{Solution properties as a function of the wedge angle parameter, $\beta$, obtained from the Falkner-Skan system~(\ref{diffeq}). (a) Dimensionless velocity gradient, $\kappa=f''(0)$, used to construct the series expansion~(\ref{series}).  (b) Asymptotic constant $B$ defined according to~(\ref{eq:farfield}).  The constants $\kappa$ and $B$ are determined numerically using the shooting method of~\cite{Cebeci} with marching done using the 4$^\mathrm{th}$-order Runge-Kutta method with a step size of $10^{-4}$.  Specific values of $(\beta, \kappa, B)$ associated with the roman numerals in the plots are as follows: I. (0.5, 0.927680039836653, $-$0.804548615), II. (0, 0.469599988361, $-$1.21678062) III. ($-$0.12, 0.28176052424, $-$1.511343148) IV. ($-$0.198837735, 0, $-$2.359), V. ($-$0.12, $-$0.1429351943576, $-$4.3989990662), VI. ($-$0.02, $-$0.065168585542904, $-$9.18639214).  Convergence  has been established to within the digits reported above by successively increasing the domain length ($\eta\in[0, L]$, $L$=10, 20, 30, 40) of the shooting method.  A tabulated version of this data is provided in Supplementary Material~\cite{supplemental}. }
\label{fig:kappa}
\end{figure}

Solutions to the Falkner-Skan equation system~(\ref{diffeq}) are found for a given flow by fixing the parameter $\beta$, which is related to the wedge angle shown in Fig.~\ref{fig:schematic}.  Note that this parameter incorporates the effect of wedge angle on the potential flow that drives the fluid motion in the boundary layer. Although~(\ref{diffeq}) is a boundary value problem, a series solution may be generated if the infinity boundary condition is relaxed, and the dimensionless velocity gradient at the wedge surface, $\kappa(\beta)$, is applied at $\eta=0$ as
\begin{equation}
f''(0)= \kappa(\beta).
\end{equation}
Note that the function $\kappa(\beta)$ is determined such that the $\eta\to\infty$ boundary condition in~(\ref{diffeq}) is applied, and is typically determined numerically. Fig.~\ref{fig:kappa}a shows the dependence of $\kappa$ on the wedge angle parameter $\beta$; the figure is constructed by implementing the shooting method of~\cite{Cebeci}.  Key values in Fig.~\ref{fig:kappa}a are as follows. When $\beta$=0 (point II in  Fig.~\ref{fig:kappa}a) there is no pressure gradient in the physical system, and the system~(\ref{diffeq}) describes the classical Blasius problem for flow along a flat plate\footnote{Note that the common form of the Blasius equation includes a coefficient of 1/2 in front of the $ff''$ term, as a result of omitting the factor of  2 in the denominator of~(\ref{eq:eta}) when deriving~(\ref{diffeq}) from the governing equations with $\beta=0$~\cite{Schlichting}.  A ramification of this is that the values reported here are $\sqrt{2}\times$ the literature value of $\kappa$=0.3320573362152 and $1/\sqrt{2}\times$ the literature value of $B$=$-$1.72078765752~\cite{Boyd2008,Lal}.}. When $\kappa$=0 (Point IV in Fig.~\ref{fig:kappa}a, $\beta\approx-0.1988$) the velocity gradient along the wedge is zero, and this condition corresponds to the onset of boundary layer separation~\cite{Schlichting}. As shown in Fig.~\ref{fig:kappa}a, there are multiple solutions to~(\ref{diffeq}) for $\beta<$0.  Negative values of $\kappa$ (e.g., Points V and VI in Fig.~\ref{fig:kappa}a) correspond to solutions of~(\ref{diffeq}) that contain a region where $f'$ is negative before eventually switching sign to tend towards the asymptotic behavior $f'\to1$, given in~(\ref{diffeq}); this is referred to as a region of ``reversed flow'' by Stewartson~\cite{Stewartson}.  

The series solution to (\ref{diffeq}) may be obtained~\cite{Sachdev} as
\begin{subequations}
\label{series}
\begin{equation}
f=\sum_{n=0}^{\infty}a_{n}{\eta}^{n},
\end{equation}
where 
\begin{equation}
a_{n+3}=\frac{\displaystyle\sum_{j=0}^{n}\beta(j+1)(n-j+1)a(j+1)a(n-j+1)-(j+1)(j+2)a_{j+2}a_{n-j}}{(n+1)(n+2)(n+3)}.
\label{recursion}
\end{equation}
\end{subequations}
The recursion above requires knowledge of the first three coefficients in order to generate the full series. The first two coefficients $a_{0} = f (0)$=0 and $a_{1} = f'(0)$=0 are given by the first two boundary conditions in~(\ref{diffeq}).  The third coefficient $a_{2}=f''(0)/2=\kappa/2$ is taken from the numerical solution described above and shown in Fig.~\ref{fig:kappa}a.  As evidenced by Figs.~\ref{fig:pointI}a through~\ref{fig:pointV}a, the utility of the series~(\ref{series}) is limited for all values of beta as it has a finite radius of convergence.  The dependence of the radius of convergence on $\beta$ is investigated in Section~(\ref{sec:complex}).

The far-field condition $f'(\infty)=1$ appearing in~(\ref{diffeq}) holds true for all wedge angles $\beta$, as does the admission of an integration constant in this limit, such that the leading-order asymptotic behavior may be written as~\cite{Coppel}
\begin{equation}
\lim_{\eta\to\infty}\left(f-\eta\right)\equiv B(\beta)
\label{eq:farfield} 
\end{equation}
where $B$\footnote{Note that the parameter $B$ is the negative of the \textit{displacement thickness} defined in terms of the similarity variables as $\delta_1=\int_0^\infty\left(1-f'\right)d\eta$~\cite{Stewartson}.  In this paper, we define the quantity $B$ such that~(\ref{eq:farfield}) is consistent with the notation typically used for the Blasius problem ($\beta=0$)~\cite{Boyd2008}.} is a function of wedge angle $\beta$; the dependence is shown in Fig.~\ref{fig:kappa}b.  Both Figs.~\ref{fig:kappa}a and~\ref{fig:kappa}b are constructed by implementing the numerical shooting method of~\cite{Cebeci},  replacing $\infty$ in~(\ref{eq:farfield}) with a finite domain length $L$, calculating $\kappa$ and $B\approx [f(\eta=L)-L]$, and recomputing the numerical solution for successively increased $L$ until $\kappa$ and $B$ are converged to within the tolerance reported in the figure caption and in~\cite{supplemental}.  Although presented in Fig.~\ref{fig:kappa} using newly generated numerical solutions, the functionality of both $\kappa$ and $B$ on $\beta$ has long been known~\cite{Stewartson}.

\section{Asymptotic Approximant \label{sec:approximant}}
\begin{figure}[!ht]
\begin{center}
\subfloat{(a) \includegraphics[width=3in]{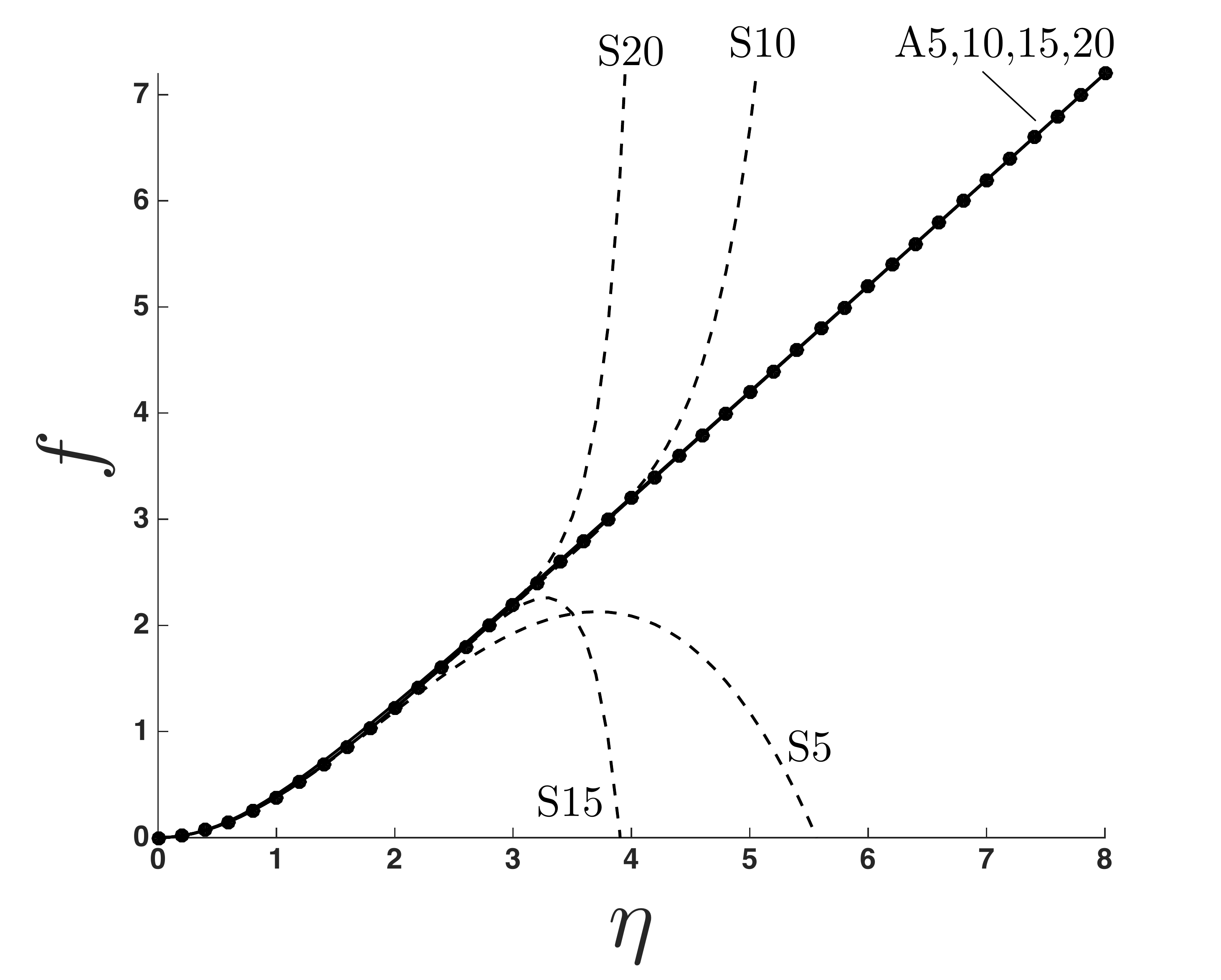}}
 \subfloat{(b) \includegraphics[width=3in]{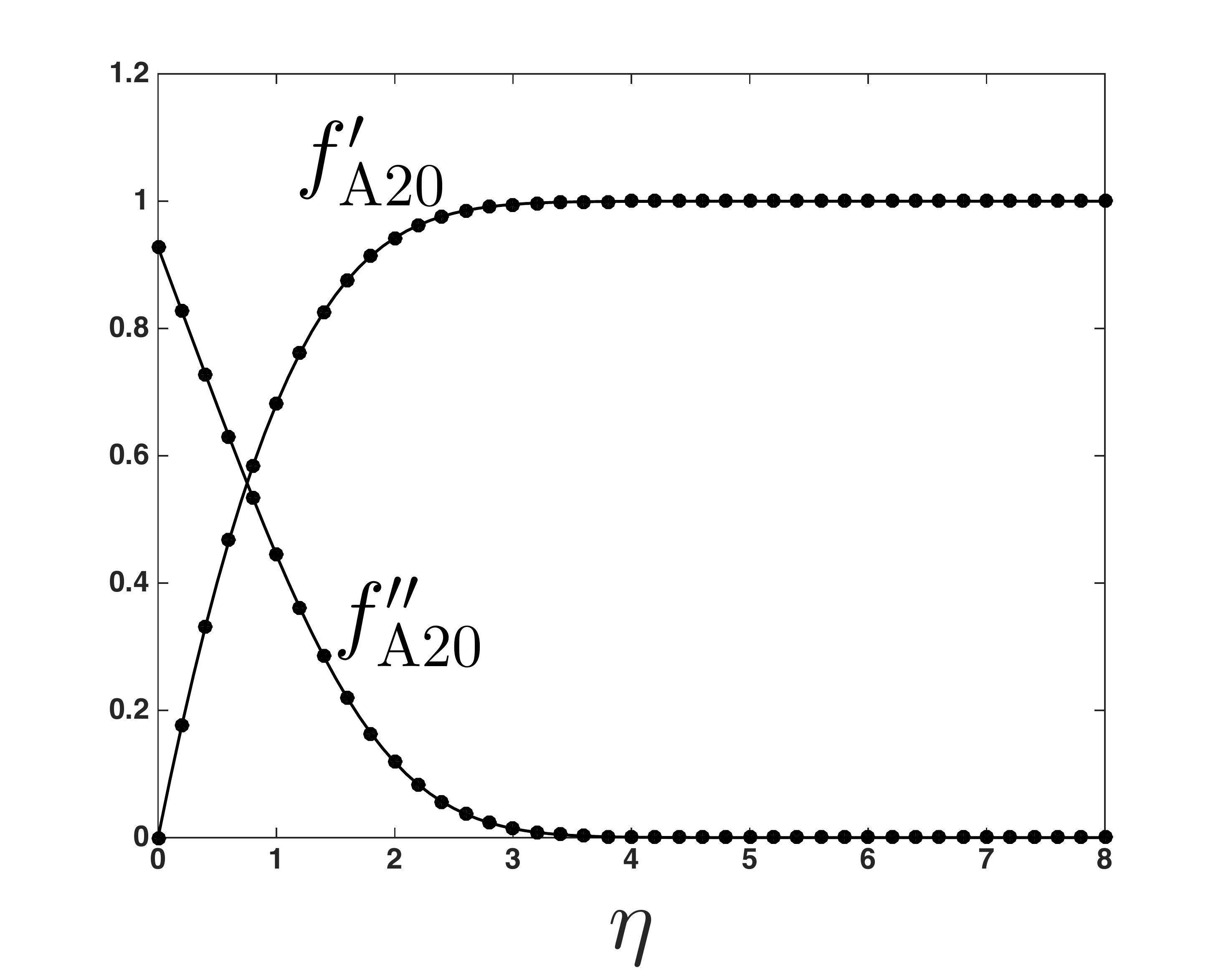}}
 \end{center}
\caption{(a) The $N$-term series~(\ref{series}) labeled S$N$ and approximant~(\ref{BlasiusA}) labeled A$N$ compared with numerical solution ($\bullet$). (b) Derivatives of approximant~(\ref{BlasiusA}) for $N$=20.   Data shown here corresponds to conditions at point I in Fig.~\ref{fig:kappa} ($\beta =0.5$, $\kappa=0.927680039836653$). }
\label{fig:pointI}
\end{figure}

\begin{figure}[!ht]
\begin{center}
\subfloat{(a) \includegraphics[width=3in]{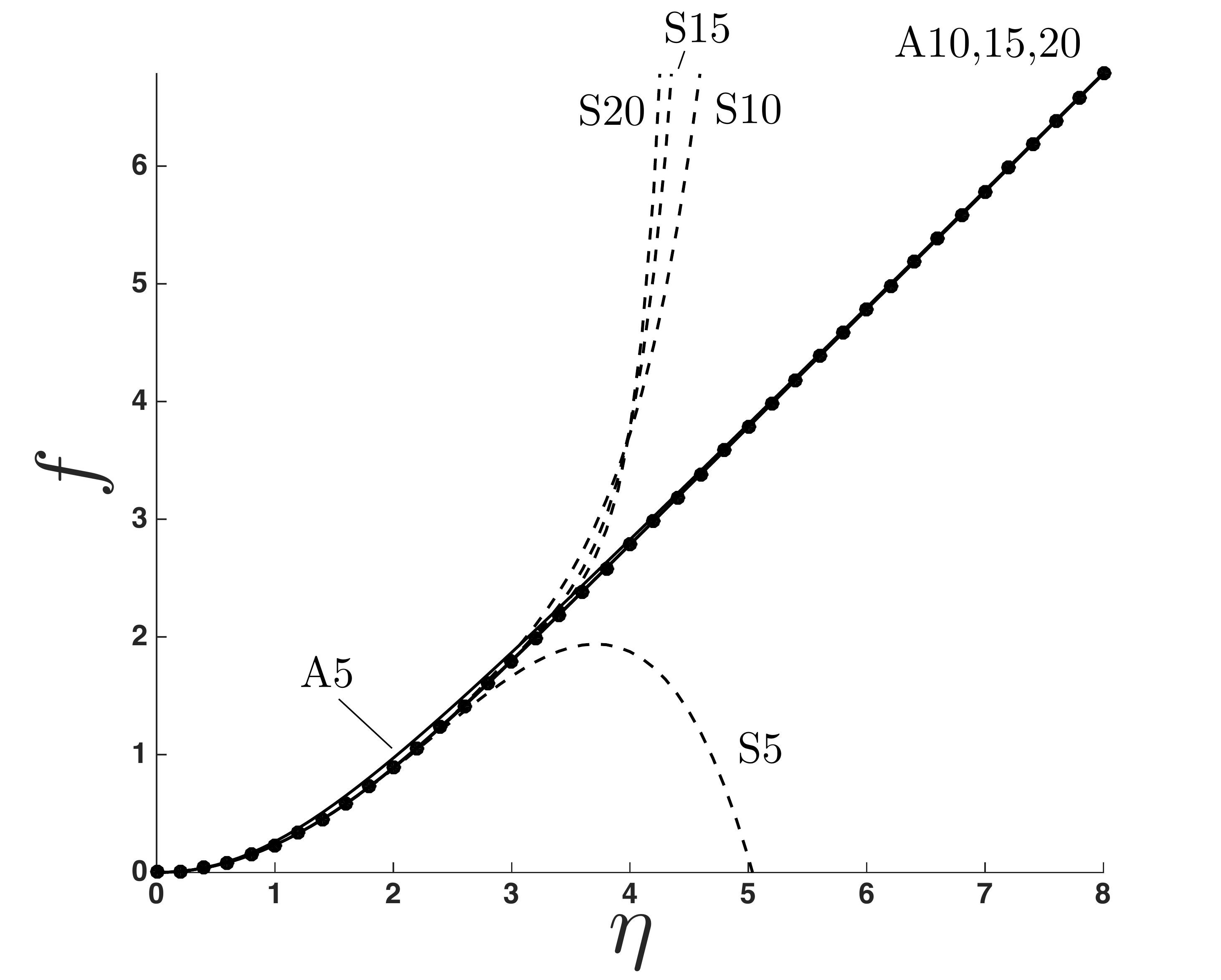}}
 \subfloat{(b) \includegraphics[width=3in]{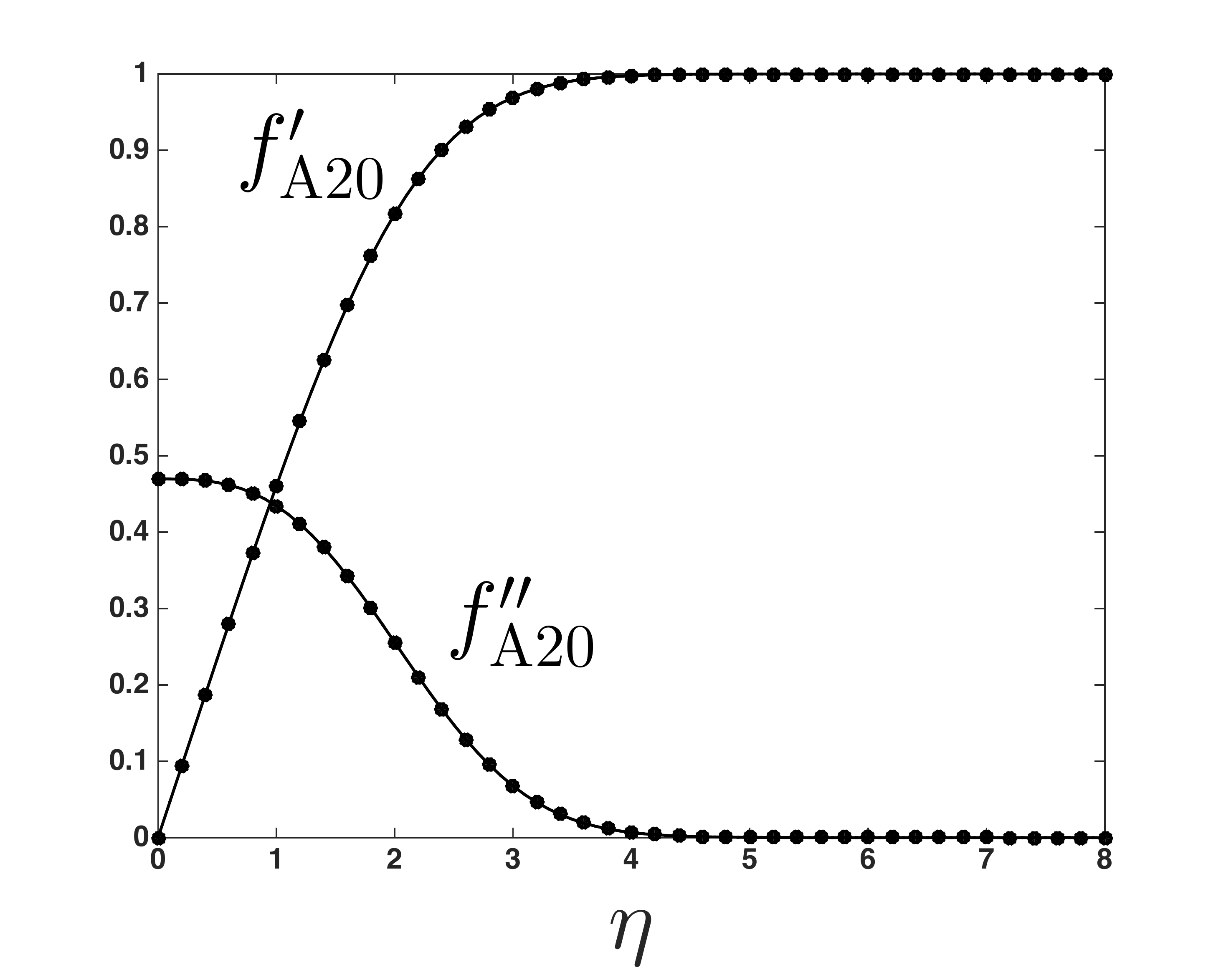}}
 \end{center}
\caption{(a) The $N$-term series~(\ref{series}) labeled S$N$ and approximant~(\ref{BlasiusA}) labeled A$N$ compared with numerical solution ($\bullet$). (b) Derivatives of approximant~(\ref{BlasiusA}) for $N$=20.   Data shown here corresponds to conditions at point II in Fig.~\ref{fig:kappa} ($\beta =0$, $\kappa=0.469599988361$). These results correspond to the classical boundary layer solution of Blasius.}
\label{fig:pointII}
\end{figure}

\begin{figure}[!ht]
\begin{center}
\subfloat{(a) \includegraphics[width=3in]{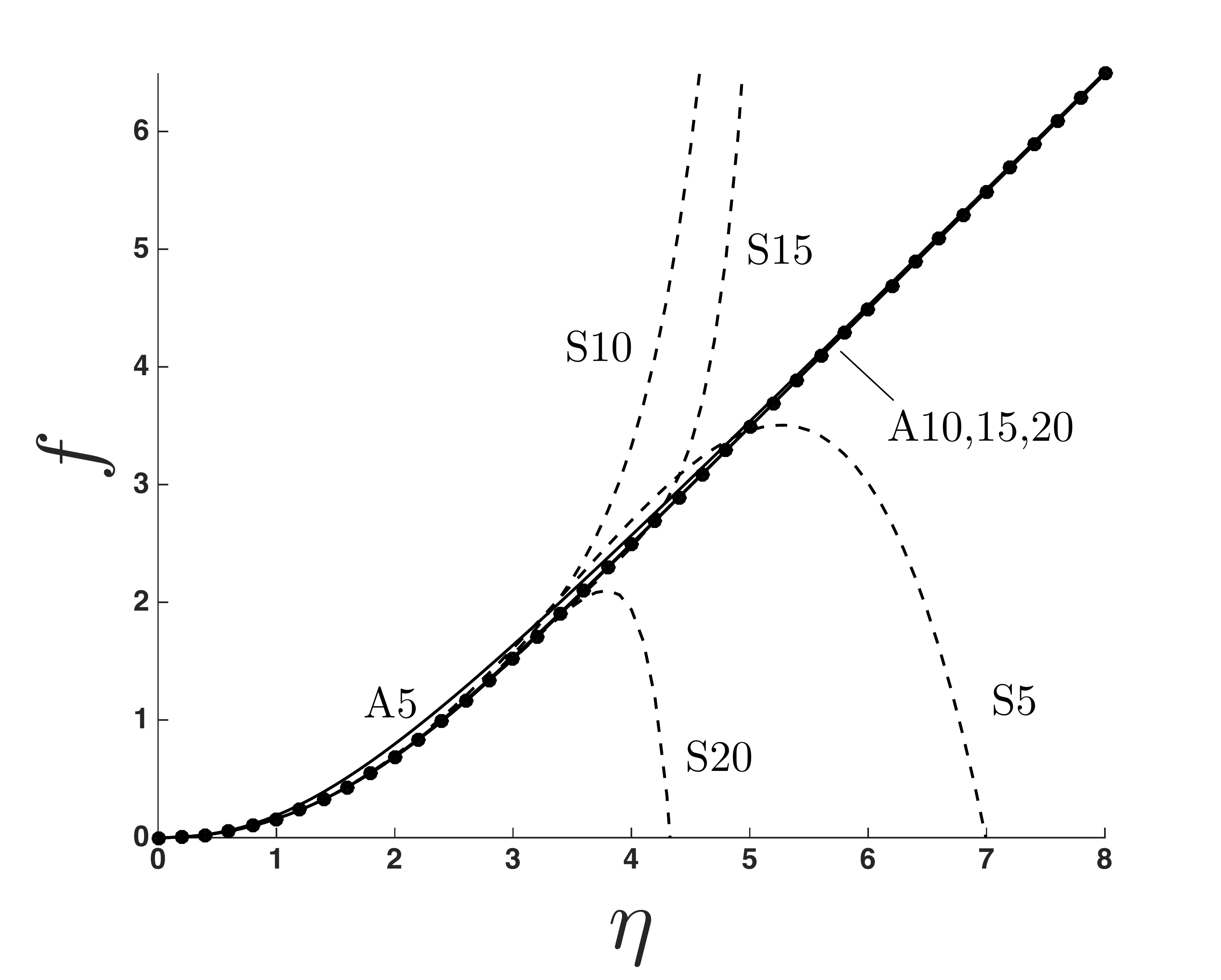}}
 \subfloat{(b) \includegraphics[width=3in]{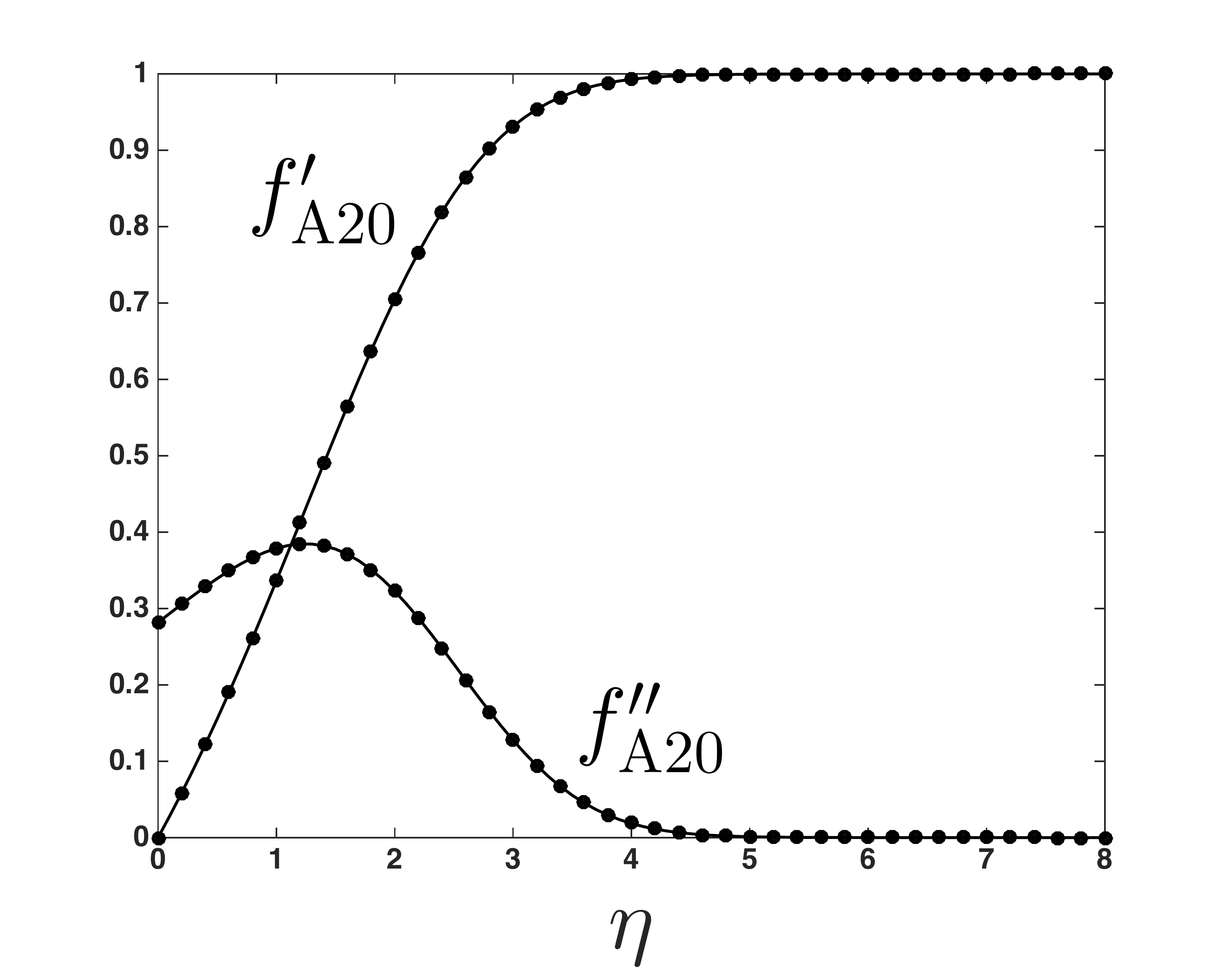}}
 \end{center}
\caption{(a) The $N$-term series~(\ref{series}) labeled S$N$ and approximant~(\ref{BlasiusA}) labeled A$N$ compared with numerical solution ($\bullet$). (b) Derivatives of approximant~(\ref{BlasiusA}) for $N$=20.   Data shown here corresponds to conditions at point III in Fig.~\ref{fig:kappa} ($\beta =-0.12$, $\kappa=0.28176052424$).  This solution is not unique \--- there is another solution for the same value of $\beta$ corresponding to point V in Fig.~\ref{fig:kappa}, as shown in Fig.~\ref{fig:pointV}.}
\label{fig:pointIII}
\end{figure}

\begin{figure}[!ht]
\begin{center}
\subfloat{(a) \includegraphics[width=3in]{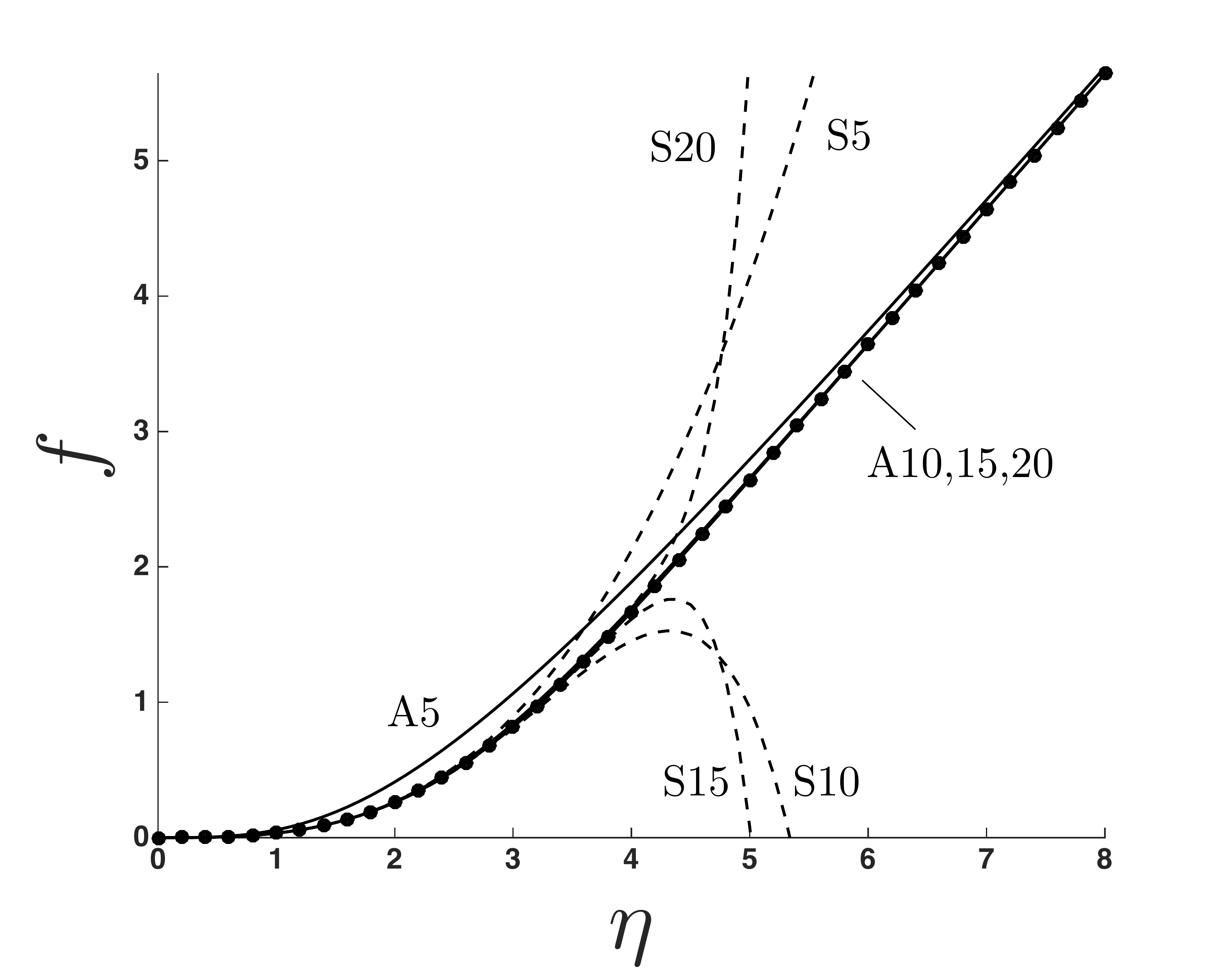}}
 \subfloat{(b) \includegraphics[width=3in]{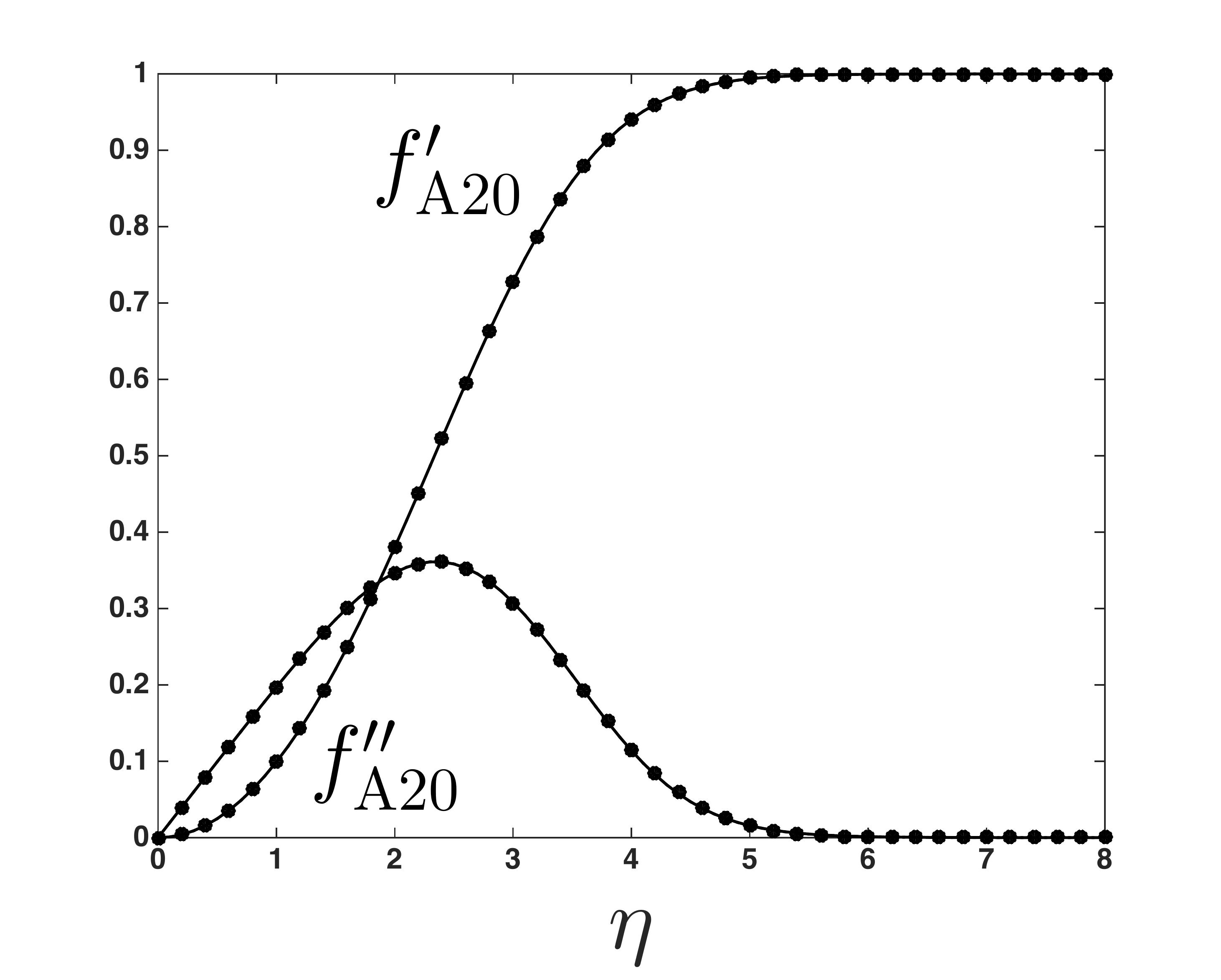}}
 \end{center}
\caption{(a) The $N$-term series~(\ref{series}) labeled S$N$ and approximant~(\ref{BlasiusA}) labeled A$N$ compared with numerical solution ($\bullet$). (b) Derivatives of approximant~(\ref{BlasiusA}) for $N$=20.   Data shown here corresponds to conditions at point IV in Fig.~\ref{fig:kappa} ($\beta =-0.198837735$, $\kappa=0$).  These results correspond to the conditions at the onset of boundary layer separation, for which the shear rate at wedge surface is zero.}
\label{fig:pointIV}
\end{figure}

\begin{figure}[!ht]
\begin{center}
\subfloat{(a) \includegraphics[width=3in]{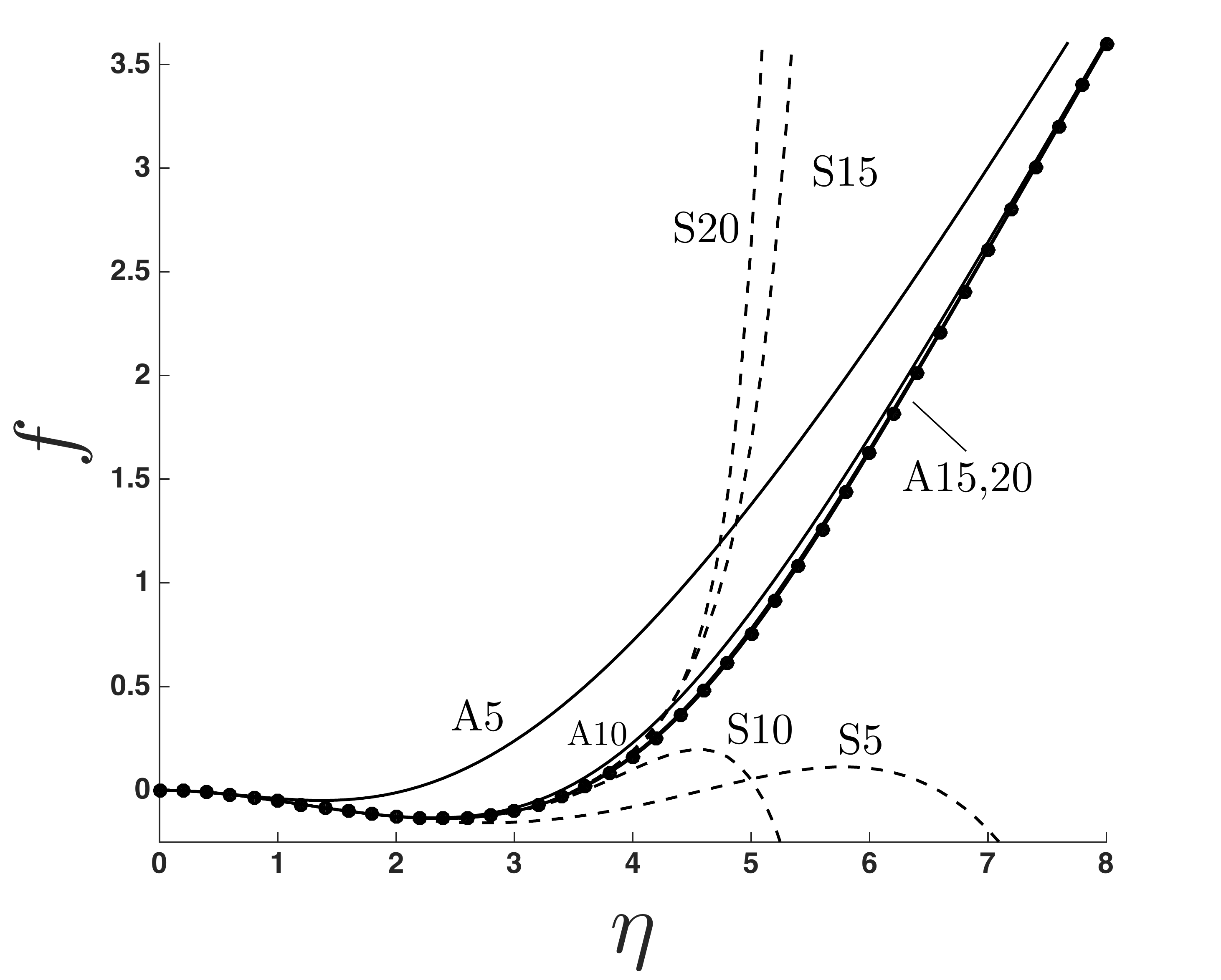}}
 \subfloat{(b) \includegraphics[width=3in]{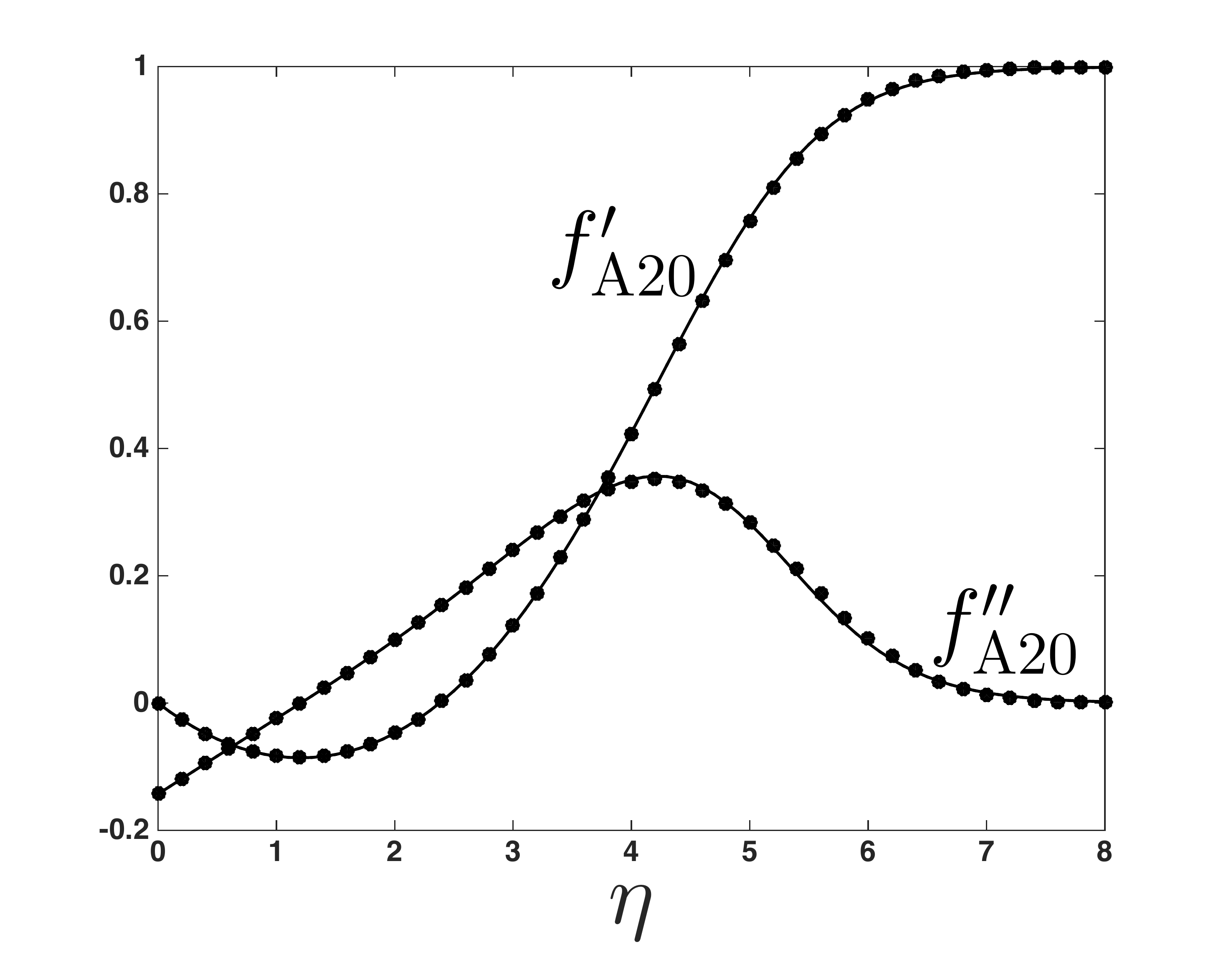}}
 \end{center}
\caption{(a) The $N$-term series~(\ref{series}) labeled S$N$ and approximant~(\ref{BlasiusA}) labeled A$N$ compared with numerical solution ($\bullet$). (b) Derivatives of approximant~(\ref{BlasiusA}) for $N$=20.   Data shown here corresponds to conditions at point V in Fig.~\ref{fig:kappa} ($\beta =-0.12$, $\kappa=-0.1429351943576$).  This solution is not unique \--- there is another solution for the same value of $\beta$ corresponding to point III in Fig.~\ref{fig:kappa}, as shown in Fig.~\ref{fig:pointIII}.}
\label{fig:pointV}
\end{figure}

\begin{figure}[!ht]
\begin{center}
 \subfloat{(a) \includegraphics[width=3in]{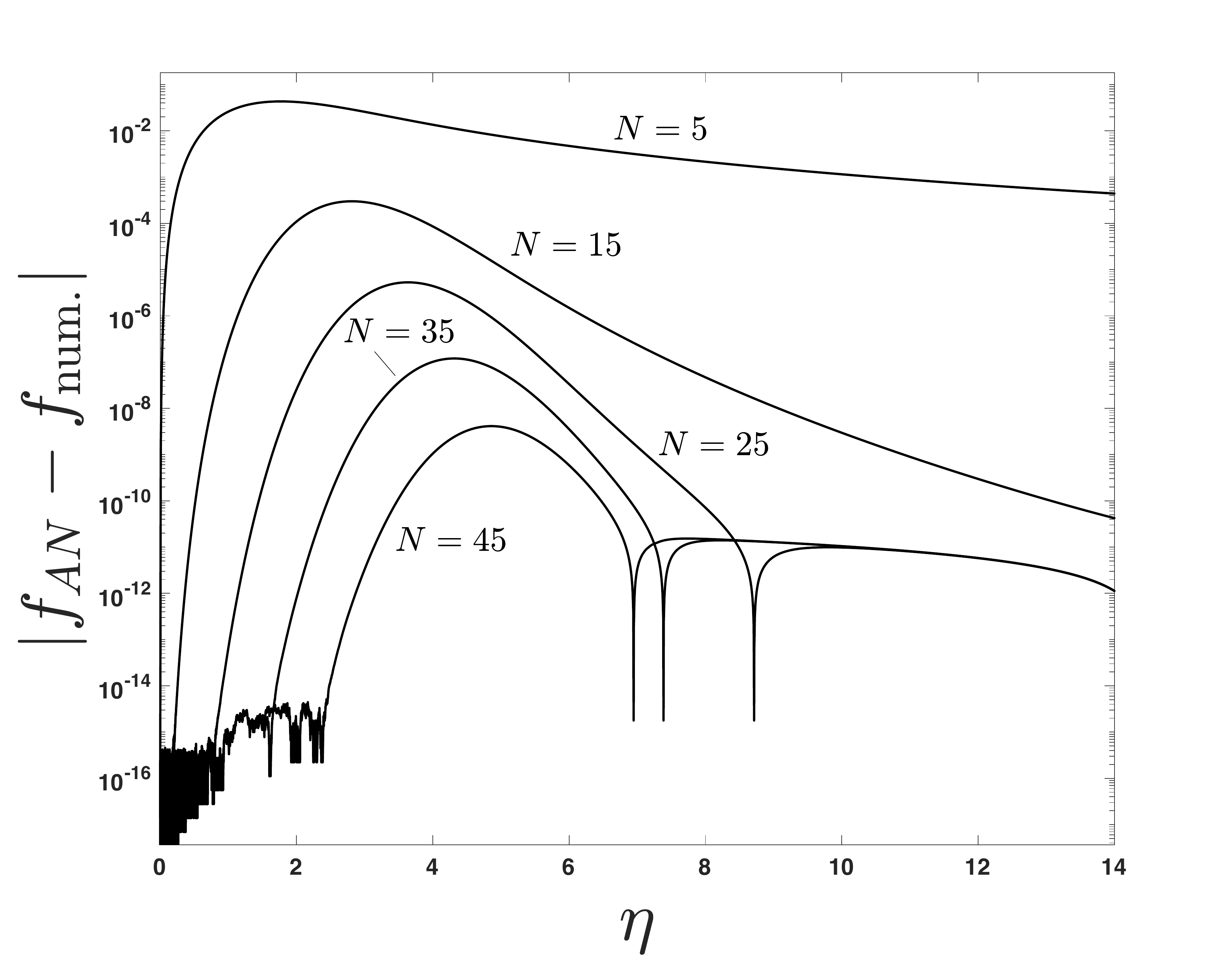}}
 \subfloat{(b) \includegraphics[width=3in]{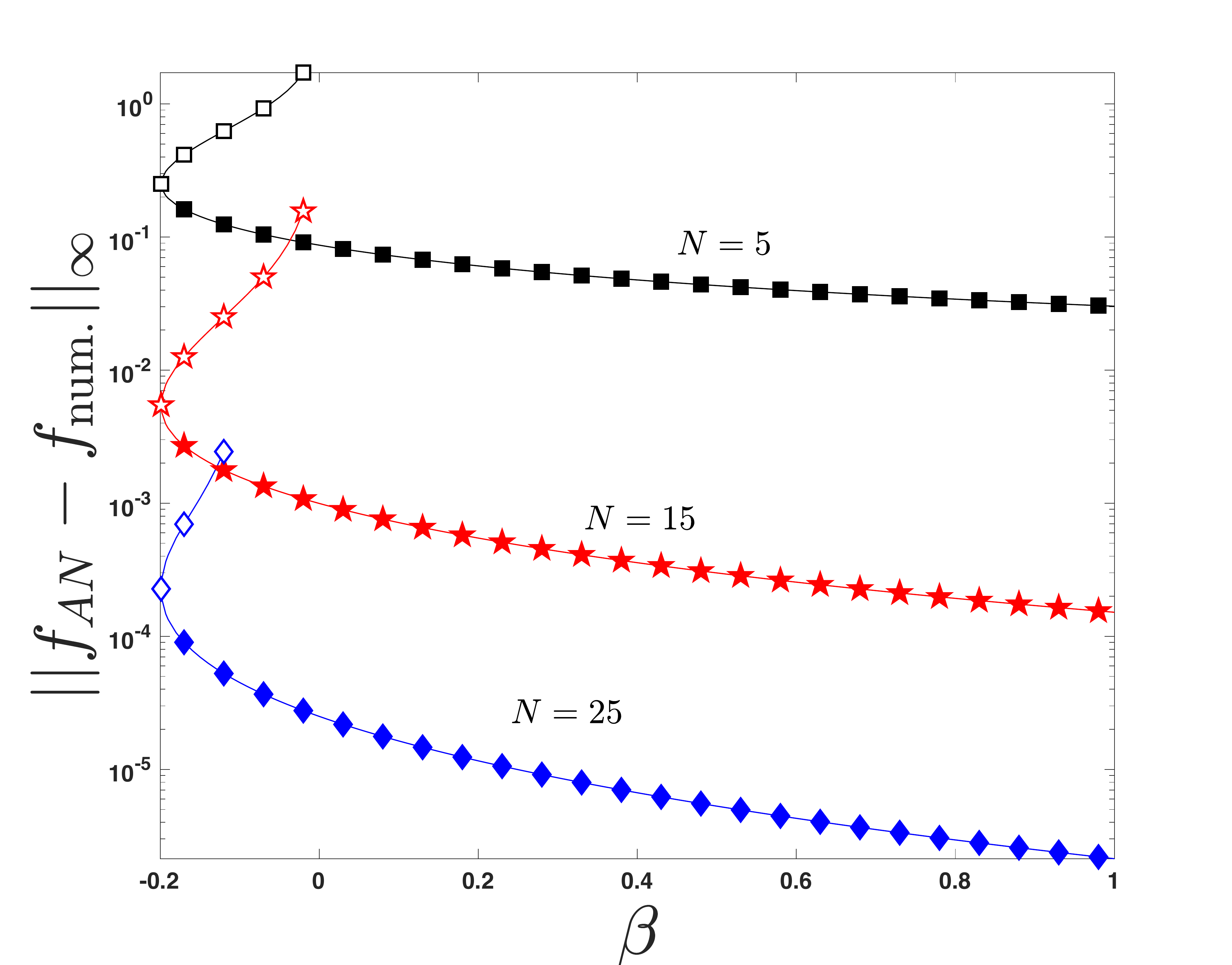}}
 \end{center}
\caption{Error associated  with approximant~(\ref{BlasiusA}), measured against numerical solutions with truncation error of O(10$^{-16}$).  (a) Absolute error in approximant~(\ref{BlasiusA}) for $\beta$=0.5. (b) Infinity norm (taken over $0\le\eta\le14$) of the absolute error associated with the approximant for $-0.198837735\le\beta\le2$.   The multivalued solutions for $\beta<0$ are represented here; the filled and open symbols correspond to  $\kappa>0$ and $\kappa<0$ solutions, respectively. 
}
\label{fig:error}
\end{figure}

\begin{figure}[!ht]
\begin{center}
\subfloat{(a) \includegraphics[width=3in]{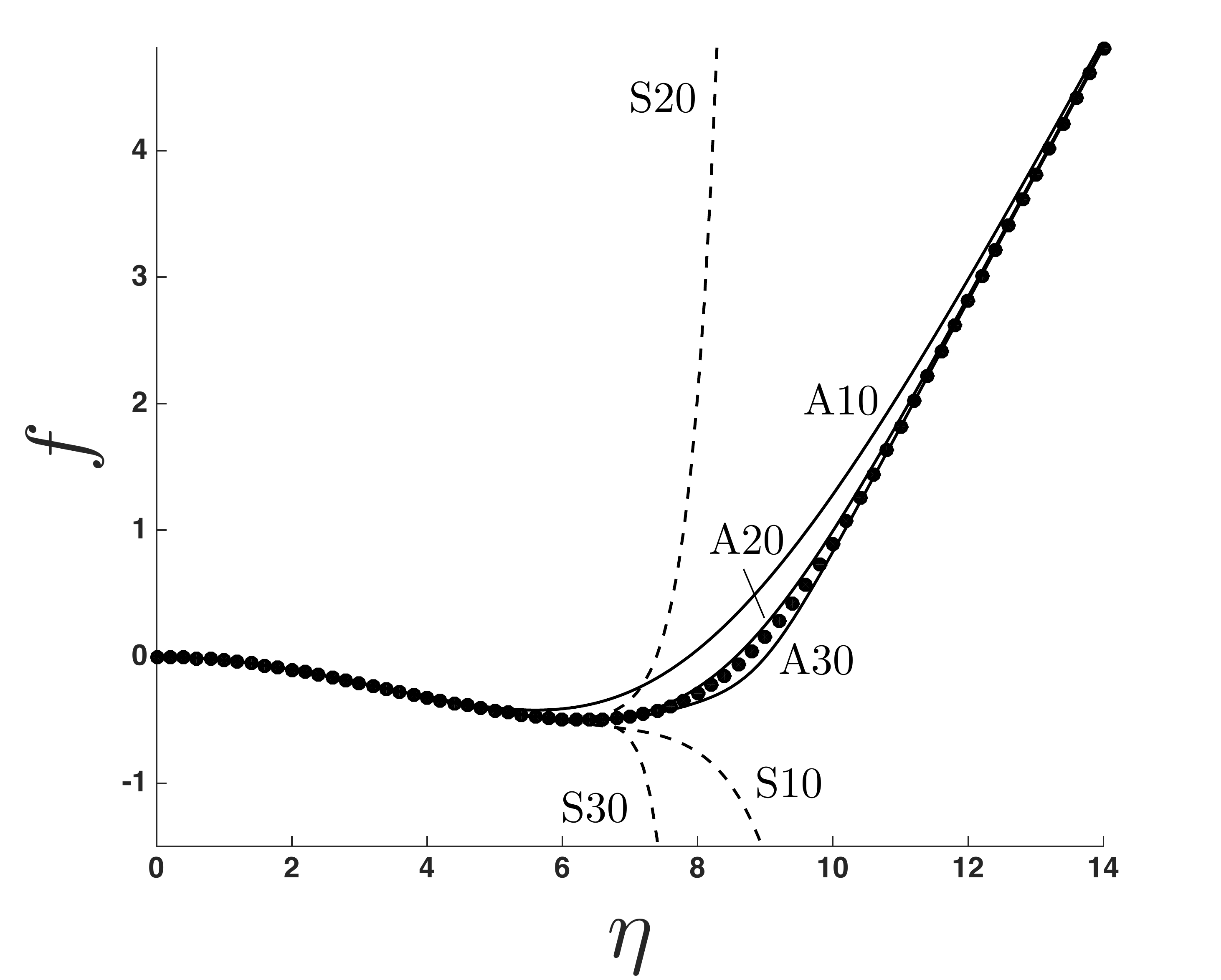}}
 \subfloat{(b) \includegraphics[width=3in]{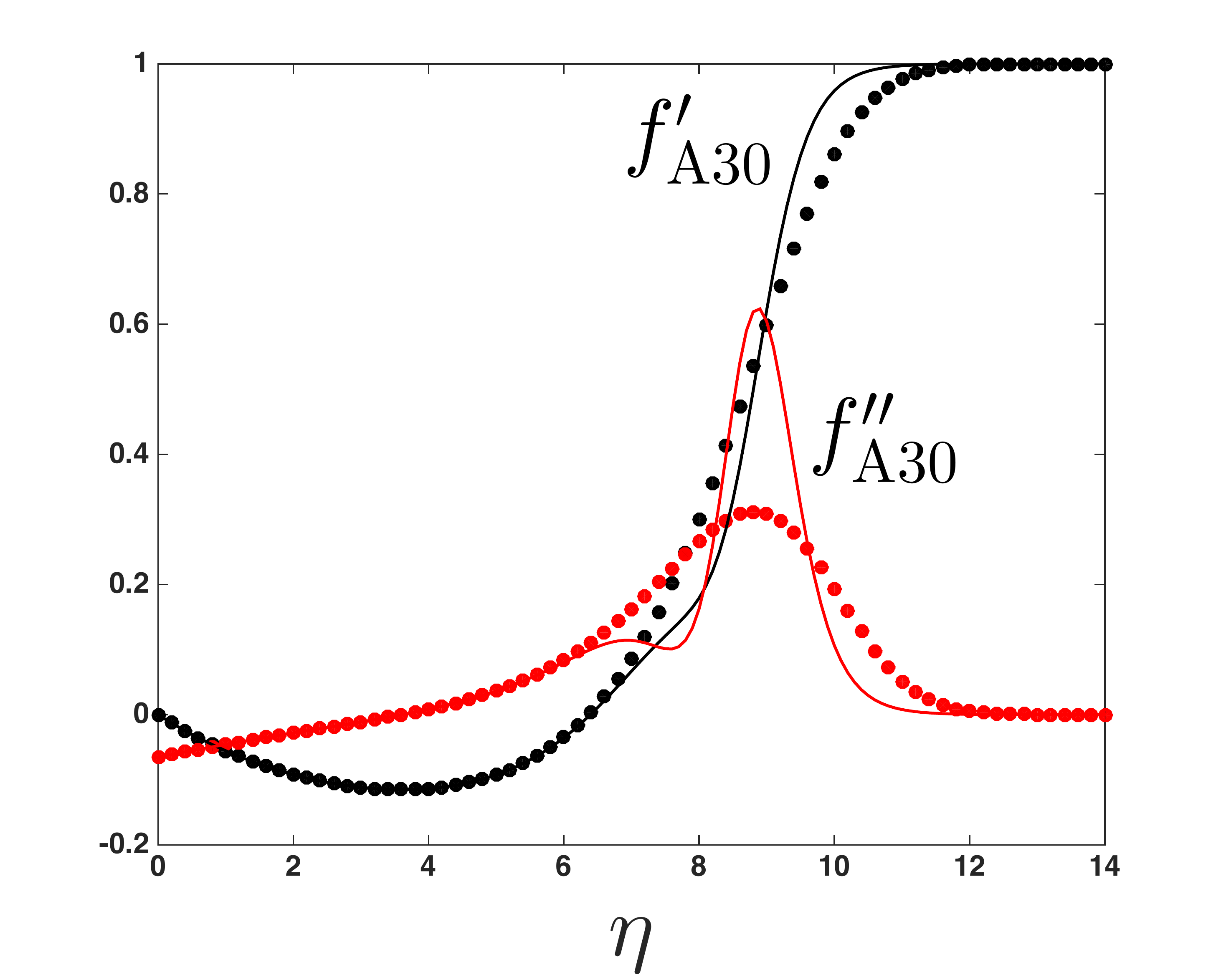}}
 \end{center}
\caption{(a) The $N$-term series~(\ref{series}) labeled S$N$ and approximant~(\ref{BlasiusA}) labeled A$N$ compared with numerical solution ($\bullet$). (b) Derivatives of approximant~(\ref{BlasiusA}) for $N$=30.   Data shown here corresponds to conditions at point VI in Fig.~\ref{fig:kappa} ($\beta =-0.02$, $\kappa=-0.065168585542904$).}
\label{fig:pointVI}
\end{figure}

The divergence of the Falkner-Skan series~(\ref{series}) demonstrated in Figs.~\ref{fig:pointI}a through~\ref{fig:pointVI}a is overcome using the method of asymptotic approximants, which constrains the analytic continuation of the series via an asymptotic behavior away from the point of expansion~\cite{Barlow:2017}.  An approximant that satisfies the $\eta\to\infty$ behavior~(\ref{eq:farfield}) is given as 
\begin{subequations}
 \label{BlasiusA}
 \begin{equation}
  f_A=\eta+B-B\left(1+\sum_{n=1}^{N}A_n~\eta^n\right)^{-1},
 \label{BlasiusApproximant}
 \end{equation}
where the $A_n$ coefficients are chosen such that the expansion of~(\ref{BlasiusApproximant}) about $\eta=0$ is exactly~(\ref{series}) up to $N^\mathrm{th}$ order.  Note that the above form is not a Pad\'e approximant, in that if one combines the terms of~(\ref{BlasiusApproximant}) through a common denominator, the coefficients of the numerator will have an explicit dependence on those in the denominator \--- this is not the case for standard Pad\'es.  The form of the approximant~(\ref{BlasiusApproximant}) is used in~\cite{Barlow:2017} for $\beta=0$ to generate accurate solutions of the Blasius boundary layer problem for flow over a flat plate.  

A recursion for the coefficients $A_n$ in~(\ref{BlasiusApproximant}) is obtained as follows. The $N$-term Taylor expansion of~(\ref{BlasiusApproximant}) about $\eta=0$ is set equal to the $N$-term truncation of~(\ref{series}).  Then, taking the limit $N\to\infty$, and re-arranging yields
\[\sum_{n=0}^\infty \tilde{a}_n\eta^n=-B\left(\sum_{n=0}^\infty A_n\eta^n\right)^{-1},\]
where (noting that $a_0=a_1=0$) $\tilde{a}_0=-B$, $\tilde{a}_1=-1$, $\tilde{a}_2=a_2=\kappa/2$, and $\tilde{a}_{j>2}=a_j$ (given by~(\ref{recursion})). Multiplying both sides of the above by the $A_n$ series and applying the well-known identity for the Cauchy product of two series~\cite{Churchill}, the expression becomes
\[\sum_{n=0}^\infty\left(\sum_{j=0}^n\tilde{a}_jA_{n-j}\right)\eta^n=-B.\]
Noting that $\tilde{a}_0=-B$, the above can be separated as
\[-BA_0+\sum_{n=1}^\infty\left(-BA_n+\sum_{j=1}^n\tilde{a}_{j}~A_{n-j}\right)\eta^n=-B.\]
Equating like terms of $\eta^0$ and $\eta^{n\neq0}$ in the expression above, we arrive at the following 
\[
A_0=1,
\]
\begin{equation}
A_{n>0}=\frac{1}{B}\sum_{j=1}^n\tilde{a}_{j}~A_{n-j},~\tilde{a}_1=-1,~\tilde{a}_{j>1}=a_j,
\label{BlasiusCoefficients}
\end{equation}
\end{subequations}
such that now all $A_n$ coefficients may be computed recursively.  Note that the result~(\ref{BlasiusCoefficients}) is identical to that provided in previous work~\cite{Barlow:2017} for the Blasius solution ($\beta=0$), but here we provide the intermediate steps for purposes of clarity.  

The approximant~(\ref{BlasiusA}) is compared with the series and numerical solutions in Figs.~\ref{fig:pointI} through~\ref{fig:pointV} for the specific values of $\beta$ enumerated as I through V in Fig.~\ref{fig:kappa} (case VI is discussed separately).  As is evident in the figures, curves for $f_{A}$ approach a final curve shape as $N$ increases.  In fact, a key property of a valid approximant is that,  upon increasing the order $N$, a convergent sequence of approximants is obtained in the sense of Cauchy. The justification for the choice of approximant form is this convergent behavior. Since all derivatives of the approximant can be computed exactly, it is not surprising that derivatives of the approximant are in excellent agreement with the numerical solution, as shown in Figs.~\ref{fig:pointI}b through~\ref{fig:pointV}b.    The advantage that the analytic form provided by the approximant has over the numerical solution is that highly-resolved domains of arbitrary length can be obtained without (a) interpolation or extrapolation of the numerical solution or (b) re-running the numerical solution at higher computational cost. 


Convergence of approximant~(\ref{BlasiusA}) is shown on a more sensitive scale in Fig.~\ref{fig:error}a for $\beta=0.5$, where the absolute error between the approximant and the numerical solution is shown for different $N$ values.   The convergence behavior shown in the figure is representative of other $\beta$ values within the range examined here.  The infinity norm of the error for $\eta\in[0, 14]$ is shown in Fig.~\ref{fig:error}b over the full range of $\beta$.  For the cases of positive $\kappa$ (filled symbols of Fig.~\ref{fig:error}b), convergence of approximant~(\ref{BlasiusA}) is established for all $\beta$, such that accurate solutions are obtained.  For the cases of negative $\kappa$ (open symbols of Fig.~\ref{fig:error}b),  convergence is apparent but is increasingly limited as $\beta\to0^-$.  Note from Fig.~\ref{fig:error}b that the overall error increases in this limit, and approximant~(\ref{BlasiusA}) becomes worse at representing the solution.  This limited convergence is demonstrated in Fig.~\ref{fig:pointVI} for $\beta=-0.02$ ($\kappa<0$, point VI in Fig.~\ref{fig:kappa}), where approximant~(\ref{BlasiusA}) is shown to poorly represent the solution for intermediate values of $\eta$; an improvement is discussed below. 

\begin{figure}[!ht]
\begin{center}
\subfloat{(a) \includegraphics[width=3in]{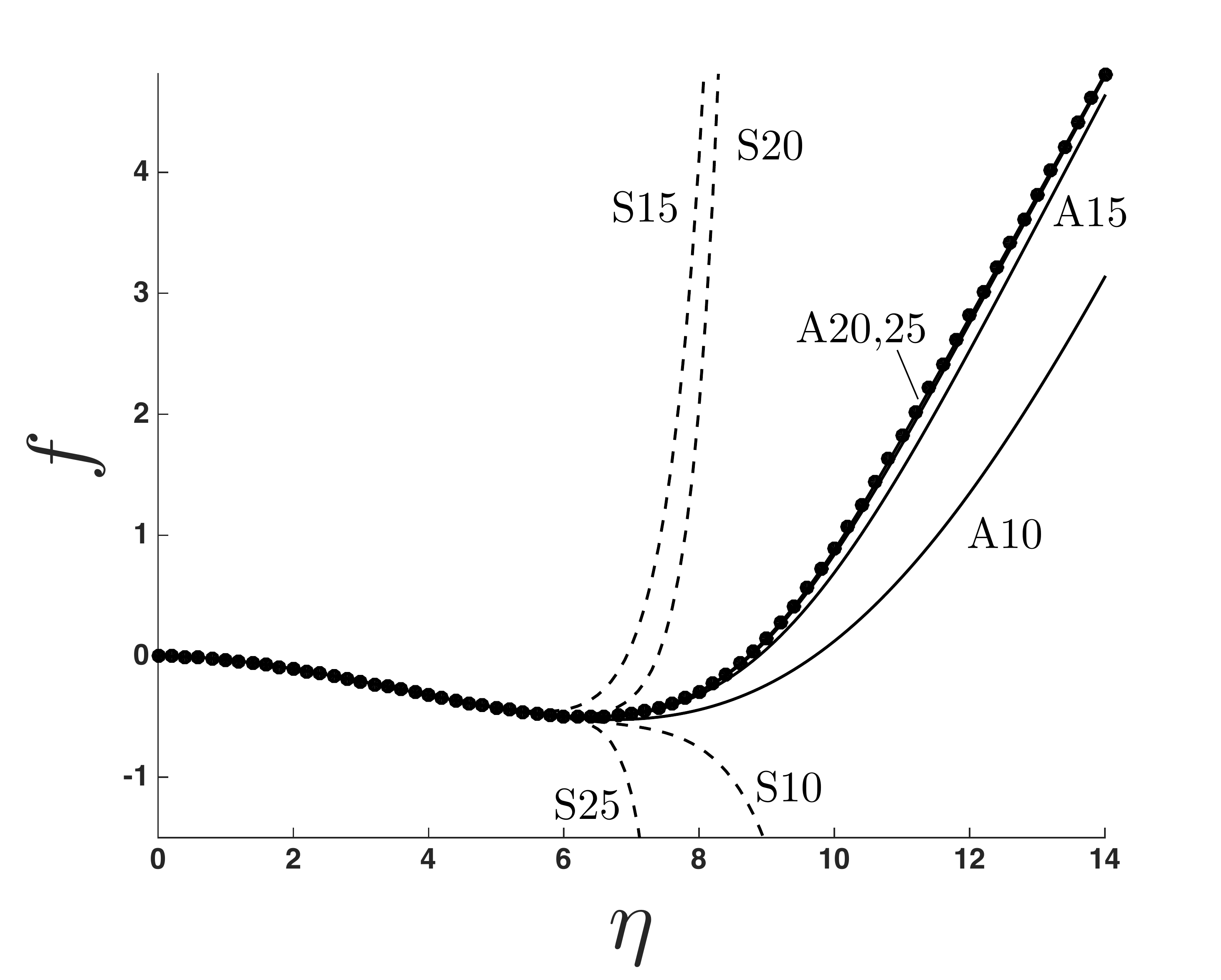}}
 \subfloat{(b) \includegraphics[width=3in]{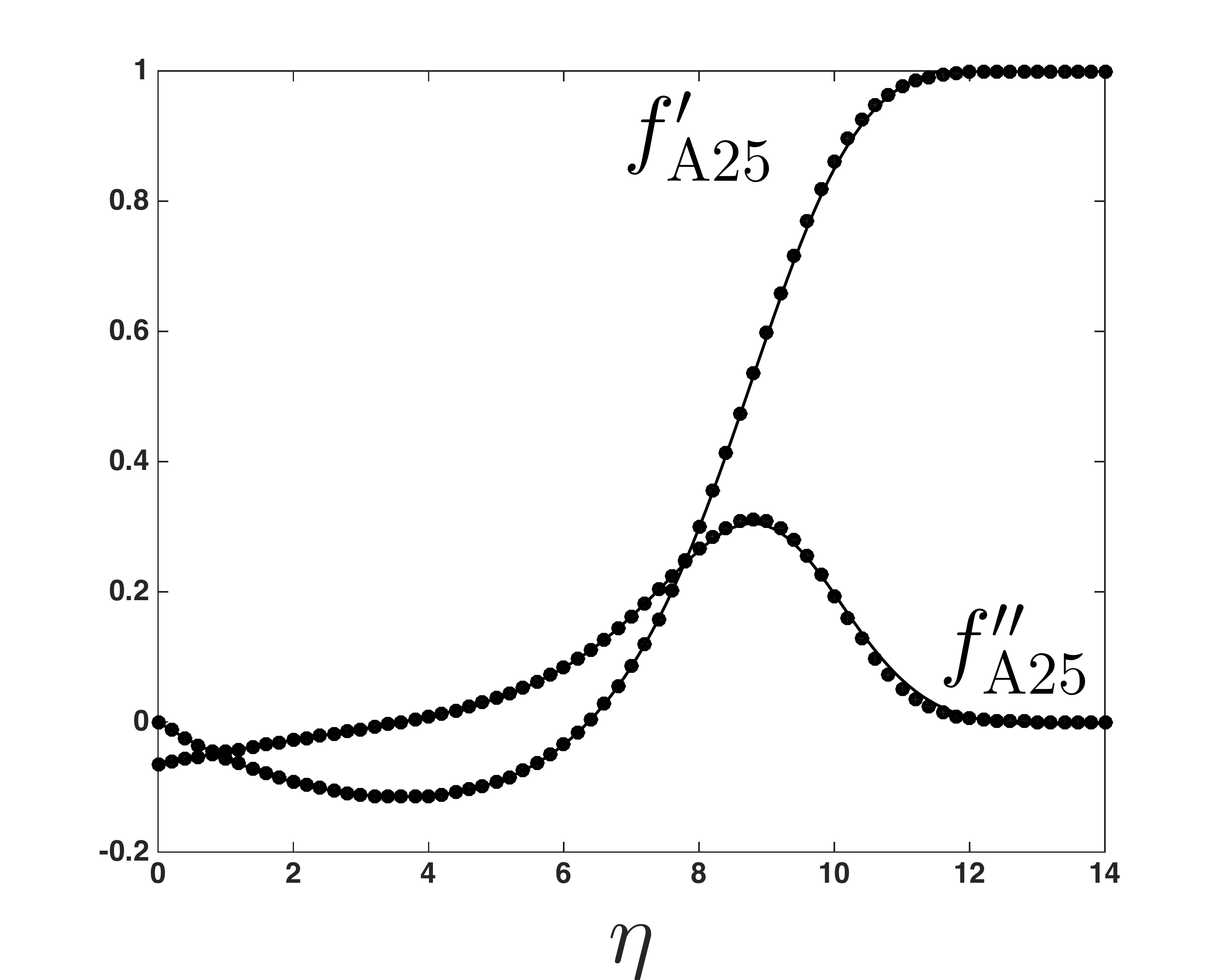}}
 \end{center}
\caption{(a) The $N$-term series~(\ref{series}) labeled S$N$ and approximant~(\ref{Approx2}) labeled A$N$ compared with numerical solution ($\bullet$). (b) Derivatives of approximant~(\ref{Approx2}) for $N$=25.   Data shown here corresponds to conditions at point VI in Fig.~\ref{fig:kappa} ($\beta =-0.02$, $\kappa=-0.065168585542904$).}
\label{fig:pointVInew}
\end{figure}

The monotonic or nearly-monotonic solutions in $f(\eta)$ shown in Figs.~\ref{fig:pointI} through~\ref{fig:pointV} (representing cases I through V in Fig.~\ref{fig:kappa}) enable the simple form of approximant~(\ref{BlasiusA}) to provide accurate solutions over the physical domain, from $\eta=0$ towards ``large'' $\eta$ such that $f'\to1$.  As $\beta$ approaches 0 from the left for $\kappa<0$, $f(\eta)$ becomes less monotonic and a minima manifests prior to reaching the $f'\to1$ large $\eta$ behavior, as shown in Fig.~\ref{fig:pointVI} for $\beta=-0.02$ (point VI in Fig.~\ref{fig:kappa}).  As mentioned above, approximant~(\ref{BlasiusA}) is incapable of correctly resolving this minima in $f(\eta)$.   To allow for more flexible curve shapes, the approximant is adjusted to have a cubic form in its numerator as follows:
\begin{equation}
f_{A}={\eta} + B - B\left[\frac{A_0 + A_1{\eta} + A_2{\eta}^2 + A_3{\eta}^3} {1+\displaystyle{\sum_{n=1}^{N-3}} d_n {\eta}^n}\right],
~~N>3.
\label{Approx2}
\end{equation}
Like~(\ref{BlasiusA}), approximant~(\ref{Approx2}) satisfies the large $\eta$ behavior~(\ref{eq:farfield}) by construction.  The unknown coefficients $A_0\dots A_3$ and $d_n$ in~(\ref{Approx2}) are chosen such that the expansion of~(\ref{Approx2}) about $\eta$=0 is equal to the series given by~(\ref{series}) to order $N$.  Note that the term in square brackets in~(\ref{Approx2}) is a [3/N-3] Pad\'e approximant, for which  solvers are readily available~\cite{Trefethen} to handle the required matrix inversion.   One may compute the $A_0\dots A_3$ and $d_n$ coefficients by isolating the bracketed term of~(\ref{Approx2}) onto one side of the equation and finding the [3/N-3] Pad\'e for the power series $1+\left(\eta-\sum_{n=0}^Na_n\eta^n\right)/B$.   As before, the constants $\kappa$ and $B$ are taken from the numerical results shown in Fig.~\ref{fig:kappa} and are inputs to the approximant. Fig.~\ref{fig:pointVInew} shows the convergence of approximant~(\ref{Approx2}) applied to $\beta=-0.02$ ($\kappa<0$, point VI in Fig.~\ref{fig:kappa}), and demonstrates a significant improvement over approximant~(\ref{BlasiusA}) (see Fig.~\ref{fig:pointVI} for comparison). Note that, although a Pad\'e solver can be used,~(\ref{Approx2}) is not a standard Pad\'e as they are defined today \--- it is a Pad\'e as it was originally intended by Baker \& Gammel~\cite{BakerGammel}, constructed such that it is consistent with the correct $\eta\to\infty$ limit.

Although an improvement is found using approximant~(\ref{Approx2}) over the simple~(\ref{BlasiusA}) for the cases of negative shear and small $\beta$, for all other $\beta$ values (including those that are physically relevant), both approximants perform equally well.  Approximant~(\ref{BlasiusA}), however, is simpler to compute, as it is based solely on recursion.   

\subsection{Complex singularities that limit convergence of power series solution~(\ref{series}) \label{sec:complex}}
\begin{figure}[!ht]
\begin{center}
\subfloat{(a) \includegraphics[width=3in]{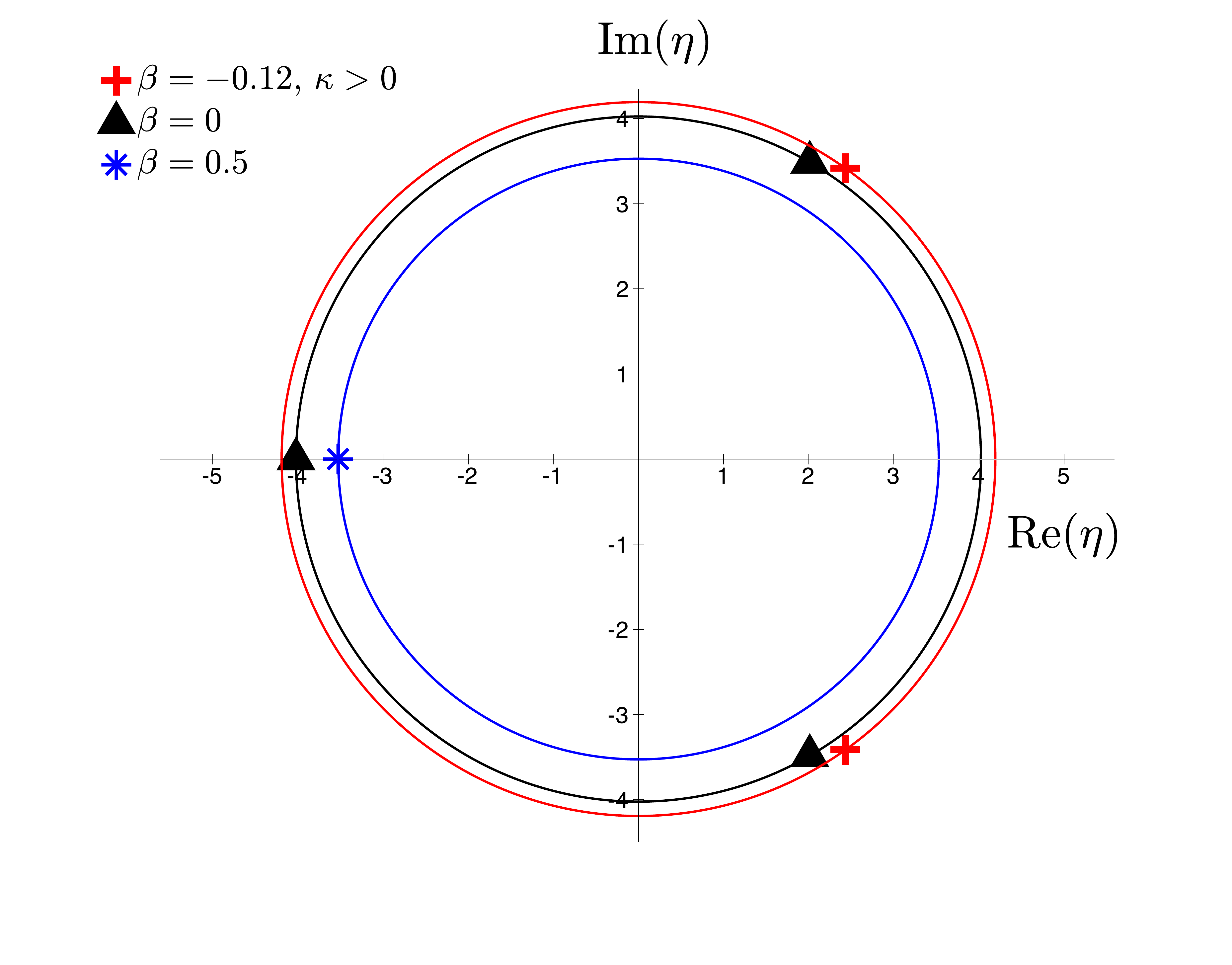}}
 \subfloat{(b) \includegraphics[width=3in]{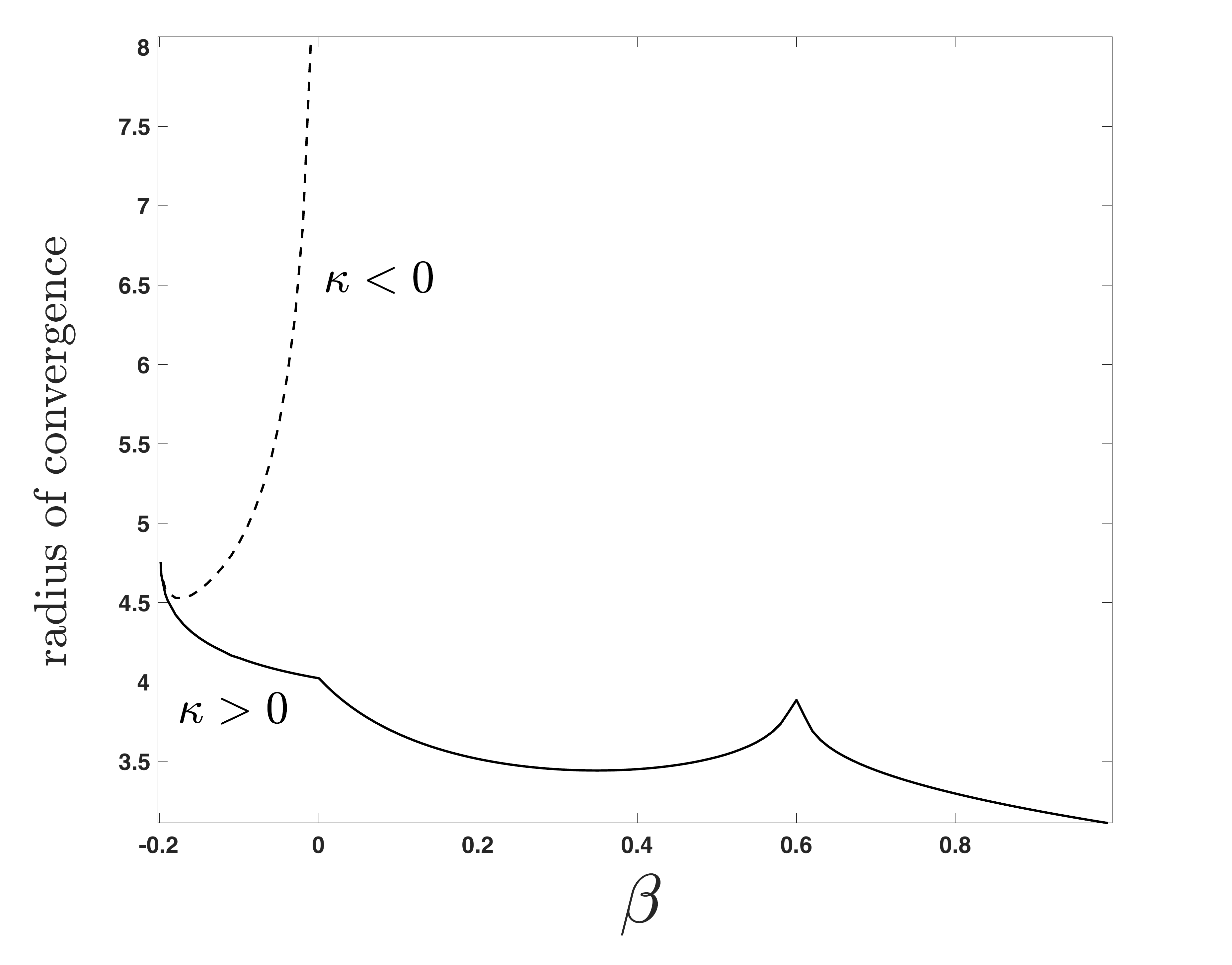}}\\
 \end{center}
\caption{(a) Location of singularities that limit convergence of the Falkner-Skan series~(\ref{series}), shown for three values of $\beta$. (b)  Radius of convergence of~(\ref{series}); a tabulated version of this curve is provided in Supplementary Material~\cite{supplemental}.}
\label{fig:sing}
\end{figure}

The power series result~(\ref{series}) is an exact representation of the solution to~(\ref{diffeq}) within its radius of convergence.  As there is no exact solution to~(\ref{diffeq}), the radius of convergence must be extracted by examining successive terms in the power series using standard methods such as the ratio test~\cite{Titchmarsh}.  Alternatively, as has been done using Pad\'e approximants~\cite{VanDyke,Guttmann2}, the asymptotic approximant~(\ref{BlasiusA}) may be used to determine the location of these singularities and the radius of convergence of the series.  The radius itself is set from the singularities closest to $\eta=0$ in the solution of~(\ref{diffeq}) when viewed in the complex plane; it follows, then, that the roots of the denominator of the approximant~(\ref{BlasiusApproximant}) should be examined.  Here, we increase the number of terms in the $A_n$ series in the denominator of~(\ref{BlasiusApproximant}) and obtain convergent solutions for the roots nearest to $\eta=0$.  Fig.~\ref{fig:sing}a provides results corresponding to three different values of $\beta$.  Roots for $\beta=0$ lie on the negative real axis, the first quadrant, and the fourth quadrant of the complex plane, and yield a radius of convergence of approximately 4.024 in agreement with Boyd~\cite{Boyd1999, Boyd2008}\footnote{As a result of using a different similarity variable, the value reported in~\cite{Boyd1999, Boyd2008} is $\sqrt{2}\times$ our value (see footnote $\dagger$ in Section~\ref{sec:series})}.  For $0<\beta<0.6$, the root structure changes, as instead only one singularity appears that sets the circle convergence, lying on the negative real axis\footnote{For $0.5<\beta<0.6$, the singularity location (orientation and radius) does not converge using~(\ref{BlasiusA}). However, the series terms eventually alternate sign at higher-order, indicating that the closest singularity lies on the negative real axis.}, as shown in Fig.~\ref{fig:sing}a for $\beta=0.5$. For $\beta<0$ ($\kappa>0$ branch) there are a pair of singularities that set circle of convergence, shown in Fig.~\ref{fig:sing}a for $\beta=-0.12$ ($\kappa>0$); this singularity orientation persists up until separation ($\beta=-0.199$). Outside of these ranges, for $\beta > -0.199$ ($\kappa < 0$) and $\beta > 0.6$, estimates for the roots using the approximant do not converge with increasing $N$. However, even when the roots do not converge, the singularities move in such a way that their distance from $\eta=0$ does converge for all $\beta$ excluding $0.5<\beta<0.6$ (over this limited range of $\beta$, the ratio-test provides an accurate radius of convergence) \--- thus, the radius of convergence of~(\ref{series}) may be determined. Fig.~\ref{fig:sing}b shows the dependence of radius of convergence on $\beta$. 
Note that, although Fig.~\ref{fig:sing}b indicates an apparent infinite radius of convergence for $\beta\to0^-$ ($\kappa<0$), Figs.~\ref{fig:pointV} and~\ref{fig:pointVI} (typical of other cases) show that the local minima in the $f(\eta)$ curve roughly tracks along with the radius of convergence and is pushed farther out as $\beta$ becomes small. Thus, although the radius of convergence increases for smaller $\beta$ past separation ($\kappa < 0$), the asymptotic behavior ($f'\to1$) is never captured by the series.

\section{Conclusions \label{sec:conclusions}}
Asymptotic approximants provide nearly exact closed-form solutions to the Falkner-Skan boundary layer equation for varying wedge angle. This adds to the increasing number of problems in disparate areas of mathematical physics to which asymptotic approximants have been applied successfully ~\cite{BarlowJCP,BarlowAIChE,Barlow2015,Barlow:2017,Barlow:2017b,Beachley}. Advantages of asymptotic approximants, specifically for the Falkner-Skan problem and in general for other problems, are their simple form, ability to yield highly accurate solutions, accuracy in solution derivatives, and low computational load. The approximant is used to determine the singularities in the complex plane that prescribe the radius of convergence of the power series solution to the Falkner-Skan equation for a large range of wedge angles.  One limitation of the methodology provided here is that accurate values of the wall velocity gradient, $\kappa$, and asymptotic parameter, $B$, are needed to construct the approximant.  As these are determined numerically in the present work, the approximants can only be utilized in post-processing once numerics are implemented.  In an approximant generated for the Sakiadis boundary layer in a previous paper~\cite{Barlow:2017}, however, it is possible to predict with high accuracy these parameters by a judicious choice in approximant coefficients.  In the same paper, the approximant used here is applied to the Blasius problem to predict $\kappa$ and $B$, but the precision does not approach that of existing benchmarks~\cite{Boyd2008} since, unlike the Sakiadis problem, higher-order asymptotic behavior is not incorporated in the approximant. Although not reported here, reasonable estimates for $\kappa$ and $B$ can be predicted applying the approximant for $0\le\beta\le1$, but the prediction method fails for $\beta<0$. That said, one may interpolate between the numerically-obtained tabulated results given in the supplementary material~\cite{supplemental} to obtain fairly accurate values for $\kappa$ and $B$ for values of $\beta$ not tabulated.  Implementation of approximants in this manner makes them fully independent of further numerical predictions.

\bibliographystyle{unsrt}
\bibliography{FS}

\begin{thebibliography}{10}
\bibitem{supplemental}
See (insert url provided by journal) for a spreadsheet of values that were
  computed in this work.
\end{thebibliography}

\end{document}